\documentclass{osa-article}

\journal{oe}

\articletype{Research Article}

\usepackage{amsmath} 
\usepackage{lineno}
\usepackage[colorlinks=true]{hyperref}

\newcommand{\bra}[1]{\left\langle #1 \right|}
\newcommand{\ket}[1]{\left|#1\right\rangle}

\newcommand{\abs}[1]{\left| #1 \right|} 
\newcommand{\iu}{{i\mkern1mu}}

\begin{document}

\title{Ab Initio Spatial Phase Retrieval via Intensity Triple Correlations}

\author{Nolan Peard,\authormark{1,2} Kartik Ayyer,\authormark{3,4} and Henry~N.~Chapman\authormark{2,4,5,6,*}}

\address{\authormark{1}Department of Applied Physics, Stanford University, Stanford, CA, USA\\
\authormark{2}Center for Free-Electron Laser Science CFEL, Deutsches Elektronen-Synchrotron DESY, Notkestr. 85, 22607 Hamburg, Germany\\
\authormark{3}Max Planck Institute for the Structure and Dynamics of Matter, 22761 Hamburg, Germany\\
\authormark{4}The Hamburg Center for Ultrafast Imaging, Universit\"{a}t Hamburg, Luruper Chausee 149, 22761 Hamburg, Germany\\
\authormark{5}Department of Physics, Universit\"{a}t Hamburg, Luruper Chausee 149, 22761 Hamburg, Germany\\
\authormark{6}Department of Physics and Astronomy, Uppsala University, Box 516, Uppsala SE-75120, Sweden}

\email{\authormark{*}henry.chapman@cfel.de} 



\begin{abstract}
Second-order intensity correlations from incoherent emitters can reveal the Fourier transform modulus of their spatial distribution, but retrieving the phase to enable completely general Fourier inversion to real space remains challenging. Phase retrieval via the third-order intensity correlations has relied on special emitter configurations which simplified an unaddressed sign problem in the computation. Without a complete treatment of this sign problem, the general case of retrieving the Fourier phase from a truly arbitrary configuration of emitters is not possible. In this paper, a general method for ab initio phase retrieval via the intensity triple correlations is described. Simulations demonstrate accurate phase retrieval for clusters of incoherent emitters which could be applied to imaging stars or fluorescent atoms and molecules. With this work, it is now finally tractable to perform Fourier inversion directly and reconstruct images of arbitrary arrays of independent emitters via far-field intensity correlations alone.
\end{abstract}

\section{Introduction}
Coherent diffractive imaging uses the stationary far-field interference of elastically-scattered light to infer the geometry of a scattering potential via Fourier analysis. Since most photodetectors perform an intensity measurement, information about the relative phases $\phi(\vec{m})$ of the scattered waves at pixels $\vec{m}$ is lost and Fourier inversion to real space is incomplete~\cite{Shechtman_2015}. This ``phase problem'' is shared across a variety of imaging modalities, including x-ray crystallography and optical microscopy, and research in each field has arrived at a variety of techniques to obtain the phase information.

Non-stationary or incoherent scattering processes are known to provide more information, as much as twice the information cut-off in an optical microscope utilising incoherent illumination or fluorescence as compared with plane-wave illumination~\cite{Goodman_2003}. The far-field intensity distribution of such a process is featureless, but the measurement of intensity-intensity correlations can nevertheless be used to extract the Fourier amplitude of the object’s structure as first demonstrated by Hanbury Brown and Twiss on the radio emission of bright stars~\cite{Brown_1956}. This approach is attractive in situations where lenses of high enough angular or spatial resolution do not exist. This is certainly the case in the X-ray regime where recent work has examined the possibility of using photon pair correlations to retrieve the Fourier spectrum of x-ray fluorescence emission~\cite{Classen_2017, Schneider_2018, Trost_2020, Ho_2021, Lohse_2021, Trost_2023}. One is still left with the phase problem, which can be solved using iterative phase retrieval~\cite{Trost_2023} when the correlations are adequately and extensively sampled by detectors with large numbers of pixels. However, it has been known since the 1960s that intensity triple correlations can reveal partial phase information directly in the form of the so-called closure phase~\cite{Twiss_1969,Gamo_1963}. This has been heavily investigated in the field of radio astronomy ~\cite{Nunez_2015, Wentz_2014, Dravins_2015, Malvimat_2014, Dravins_2013} with the aim to develop the means to reconstruct an image of an arbitrary arrangement of emitters without the use of additional constraints.


Retrieval of the Fourier phase, $\phi(\vec{m})$, begins by first computing the absolute value of the \emph{closure phase}, $\abs{\Phi(\vec{m},\vec{n})} = \abs{\phi(\vec{m}+ \vec{n}) - \phi(\vec{m}) - \phi(\vec{n}) }$, from the triple correlations. Unfortunately, we need the signed value of $\Phi$ to recover completely $\phi(\vec{m})$ for the following reason: At the start of phase extraction, an estimate value is chosen for the first pixel. But for the next pixel, the sign ambiguity of $\abs{\Phi}$ returns two possible values of $\phi(\vec{m}+1)$ and an additional two possible values for every subsequent pixel (except for special arrays where $\text{sgn}(\Phi)$ may be assumed constant). To avoid this exponential expansion of the solution space, we show how redundant information contained in $\abs{\Phi}$ may be used to constrain the possible values of $\text{sgn}(\Phi)$. Multiple publications have described the concept but, to the best of our knowledge, no one has yet provided complete or useful details on calculating $\text{sgn}(\Phi(\vec{m},\vec{n}))$~\cite{Sato_1978, Sato_1979, Sato_1981, Fontana_1983, Lohmann_1983, Lohmann_1984, Bartelt_1984, Matsuoka_1984, Baldwin_1986, Yellott_1992, Marathay_1994, Sayrol_1995, Holambe_1996, Shoulga_2017}. Ab-initio phase retrieval from the third-order intensity correlations has thus remained incomplete for decades. With our method, it is now possible to solve for the phase of an arbitrary array of incoherent emitters from the third-order intensity correlations alone. Combined with the second-order intensity correlations, we have a completely general method for reconstructing images of arrays of incoherent emitters.

In this paper, we describe our solution to the sign problem of the closure phase, with the help of a simple 1D example using round numbers in section~\ref{sect:1D_Example}, and show a numerical implementation of our method with simulated data from classical independent light sources. This same method may be used to reconstruct images of star clusters or, with some corrections to account for the use of a quantum light source, arrays of fluorescent molecules or atoms. 

\begin{figure}[htb]
\centering\includegraphics[width=\textwidth]{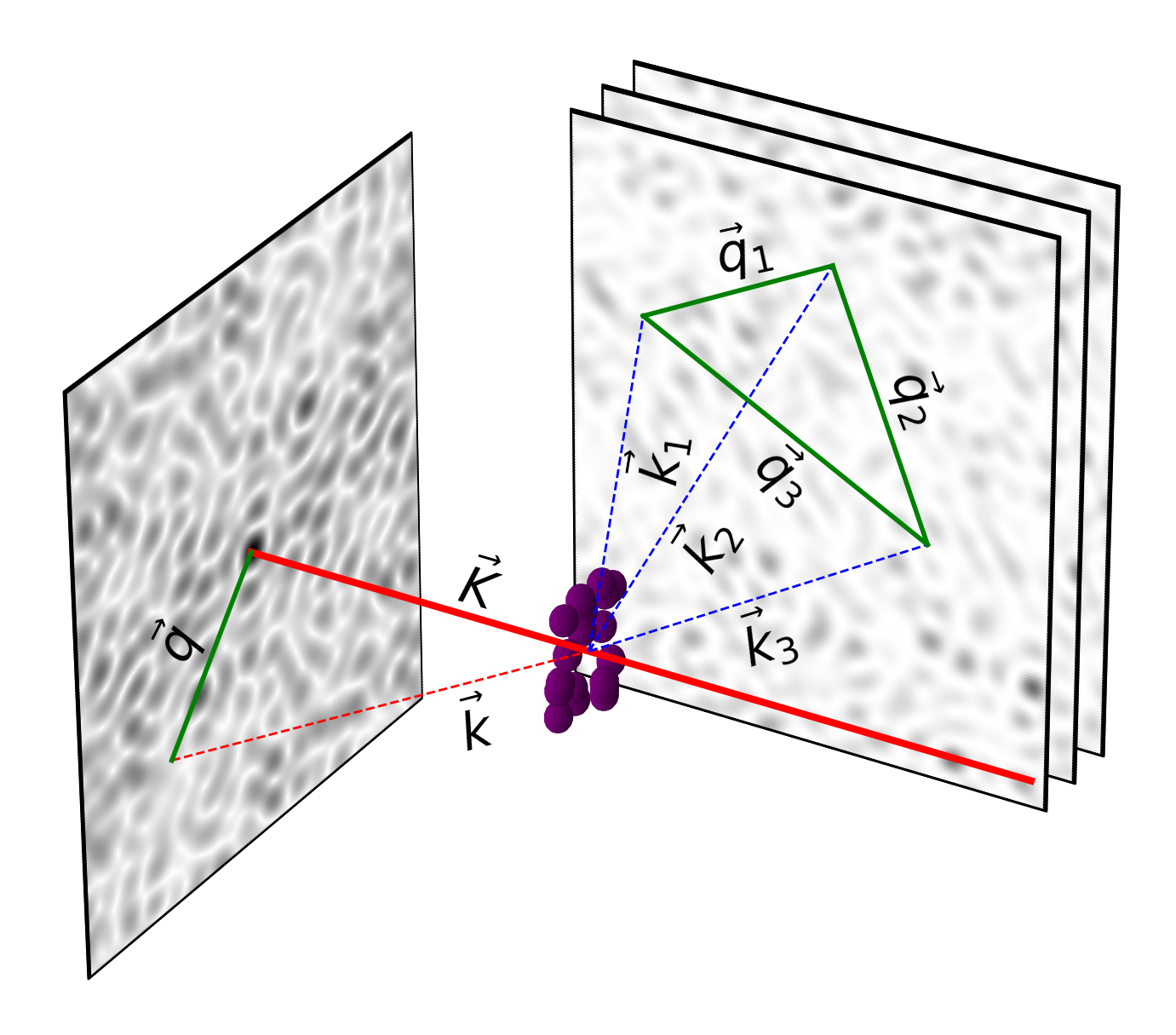}
\caption{Schematic sketch of coherent diffraction in the forward detection plane, intersecting with the incident beam $\vec{K}$. Fluorescence speckle is emitted by the atoms isotropically and the position of the second detector is not dependent on the incident beam. Coherent diffraction data is collected as a function of the scattering vector $\vec{q} = \vec{k}-\vec{K}$. Correlations between triples of fluorescence photons (intensities) at pixels separated by $\vec{q}_1$, $\vec{q}_2$, and $\vec{q}_3$ reveal the spatial phase information lost in the coherent diffraction experiment.
\label{fig:Intro}}
\end{figure}

\section{Theory}
The diagram in Fig.~\ref{fig:Intro} depicts and contrasts structure determination via coherent scattering to that obtained from incoherent emission. When illuminated with a plane wave with a wave-vector $\vec{K}$, the elastically scattered field has stationary intensity given by
\begin{equation}
    I(\vec{q}) = \left| \sum_i^\nu f_i e^{i \vec{q} \cdot \vec{r}_i }\right|^2 = \sum_{ij} f_i f_j^*  e^{i \vec{q}\cdot (\vec{r}_i-\vec{r}_j)}
    \label{eq:coherent}
\end{equation}
for a number $\nu$ of point scatterers with scattering factors $f_i$ and positions $\vec{r}_i$ relative to an arbitrary real-space origin. The photon momentum transfer $\vec{q}$ is equal to the difference $\vec{k}-\vec{K}$. The phase of each scattered wave, $\vec{q} \cdot \vec{r}_i$, is derived from the difference in the optical path along the directions of the incoming and outgoing waves as compared to a scatterer at the origin.  The intensity pattern is thus proportional to the square modulus of the Fourier transform $F(\vec{q})$ of the distribution of scatterers in terms of spatial frequencies equated with $\vec{q}$. The origin of the pattern, $\vec{q}=0$, is located in the direction of the incident beam. In this forward direction all scattered waves are in phase and there is strong constructive interference, with intensity generally falling with scattering angle. The pattern consists of speckles whose width is inversely proportional to the extent of the object. The recovery of the object's scattering potential is obtained by an inverse Fourier transform of $F(\vec{q})$, but only after the corresponding phases are obtained.

If, instead, the object consists of a collection of incoherent point emitters, then there is no dependence on any incident beam and the phase of the emission, relative to that of an emitter at an arbitrary real-space origin, is $\vec{k} \cdot \vec{r}_i + \phi_i$. We assume that the emission phases $\phi_i$ are random and uncorrelated on timescales greater than the relevant system coherence time, $\tau_c$, due to independent, spontaneous emission at random times.   
The total light field in this scenario is often referred to as pseudo-thermal or chaotic and has intensity
 \begin{equation} 
   I(\vec{k}) = \left| \sum_i^\nu s_i e^{i(\vec{k} \cdot \vec{r}_i + \phi_i(t>\tau_c))} \right|^2 = \sum_{ij} s_i s_j^* e^{i\left(\phi_i(t>\tau_c)-\phi_j(t>\tau_c)\right)} e^{i\vec{k} \cdot (\vec{r}_i-\vec{r}_j)} 
   \label{eq:incoherent}
 \end{equation} 
with $s_i$ the amplitude of electric field emission of the $i$th emitter. The intensity pattern depends on the orientation of the object, and, given the complete independence of emission, at an instant of time this pattern has a uniform intensity modulated by speckles of the same size as the case for coherent scattering. When rapid exposures are measured with a photodetector, we can consider the phases $\phi_i$ are reset shot-to-shot, changing the instantaneous speckle pattern. From the right-hand side of Eqn.~\ref{eq:incoherent}, we observe that the structure (sum of $\vec{r}_i$ for all $i$) would be difficult to discern by averaging intensities over many shots---the random phase resets would drive the interference speckle visibility to zero. However, it remains possible to obtain structural information via intensity correlations~\cite{Classen_2017}.

In the following, we use the word \textit{atom} to refer to any member of a collection of point fluorescent (atoms and molecules) or thermal (stars) light sources. We assume these atoms to undergo spontaneous emission independently, i.e. that each atom emits a field with a phase or time delay that is uncorrelated to the fields emitted by the other atoms.

\subsection{Intensity Correlations}
We consider photon emission vectors in reciprocal space, $\vec{k}_1$, $\vec{k}_2$, and $\vec{k}_3$, and their vector differences
 \begin{align} & \vec{q}_1 = \vec{k}_1 - \vec{k}_2 \\  & \vec{q}_2 = \vec{k}_2 - \vec{k}_3 \\  & \vec{q}_3 = \vec{k}_3 - \vec{k}_1 = -\vec{q}_1 - \vec{q}_2 
 \end{align} 
as depicted in Fig.~\ref{fig:Intro}. The ensemble average of third-order intensity correlations of the light field, Eqn.~\ref{eq:incoherent}, over all shots 
 \begin{equation}
     \left\{ g^{(3)}(\vec{k}_1, \vec{k}_2, \vec{k}_3) \right\} = \left\{ \frac{\langle I(\vec{k}_1) I(\vec{k}_2) I(\vec{k}_3) \rangle }{\langle I(\vec{k}_1) \rangle  \langle I(\vec{k}_2) \rangle  \langle I(\vec{k}_3) \rangle } \right\}
 \end{equation}
can be expressed as 
 \begin{subequations}
 \begin{align} 
  \left\{ g^{(3)}(\vec{q}_1, \vec{q}_2) \right\} \approx  \left(1-\frac{3}{\nu}+\frac{4}{\nu^2}\right) & +  \left(1-\frac{2}{\nu}\right)\left(|g^{(1)}(\vec{q}_1)|^2+|g^{(1)}(\vec{q}_2)|^2+|g^{(1)}(-\vec{q}_1-\vec{q}_2)|^2\right)\\ &+ 2\text{Re}\left(g^{(1)}(\vec{q}_1) \quad g^{(1)}(\vec{q}_2) \quad g^{(1)}(-\vec{q}_1-\vec{q}_2)\right) \label{eq:ClosureTerm}
 \end{align}\label{eq:Bispectrum}
 \end{subequations}
and is called the \emph{bispectrum}. Similarly, the mean second-order intensity correlation function
 \begin{equation}
     \left\{ g^{(2)}(\vec{k}_1, \vec{k}_2) \right\} = \left\{ \frac{\langle I(\vec{k}_1) I(\vec{k}_2) \rangle }{\langle I(\vec{k}_1) \rangle  \langle I(\vec{k}_2) \rangle  } \right\}
 \end{equation}
may be written as 
 \begin{equation} 
   \left\{ g^{(2)}(\vec{q}_1) \right\} \approx 1 -\frac{1}{\nu} + \left | g^{(1)}(\vec{q}_1) \right |^2 \label{eq:SiegertModified} 
 \end{equation}
This equation is often referred to as the Siegert Relation in quantum optics~\cite{Loudon_2000}. For a full derivation of Equations \ref{eq:Bispectrum} and \ref{eq:SiegertModified} please review the Supplement.

In Eq.~\ref{eq:Bispectrum}, we have an expression for $g^{(3)}$ in terms of constants, the square modulus of $g^{(1)}$, and the real part of a product of complex-valued $g^{(1)}$. Since $\abs{g^{(1)}}$ may be acquired from $g^{(2)}$ in Eq.~\ref{eq:SiegertModified}, it is possible to extract the last term \ref{eq:ClosureTerm} alone. This term is referred to as the \emph{closure} in the astronomy literature. We can rewrite the \emph{closure} as 
 \begin{multline} 
     2\text{Re} \left(g^{(1)}(\vec{q}_1) \quad g^{(1)}(\vec{q}_2) \quad g^{(1)}(-\vec{q}_1-\vec{q}_2)\right) = \\  2\abs{g^{(1)}(\vec{q}_1)}\abs{g^{(1)}(\vec{q}_2)}\abs{g^{(1)}(-\vec{q}_1-\vec{q}_2)} \cos\Big(\phi(\vec{q}_1)+\phi(\vec{q}_2)+\phi(-\vec{q}_1-\vec{q}_2)\Big) \label{eq:Closure} 
 \end{multline} 
where we have expressed $g^{(1)}$ in polar coordinates in the complex plane. As the radial component ($\abs{g^{(1)}}$) is easily obtained from $g^{(2)}$, the phase information, $\phi(\vec{q})$, can be isolated as follows.

Suppose we set $\vec{q}_1 = \vec{m}$ and $\vec{q}_2 = \vec{n}$ where $\vec{m}$, $\vec{n}$ map to discrete pixels on a detector. The symmetry of $\abs{g^{(1)}(\vec{q})}$ and anti-symmetry of $\phi(\vec{q})$ allow us to rearrange the \emph{closure} into 
 \begin{multline}
    \cos \Big(\phi(\vec{m}+\vec{n})-\phi(\vec{m})-\phi(\vec{n})\Big) \approx \\ \frac{g^{(3)}(\vec{m},\vec{n}) - (1-\frac{3}{\nu} + \frac{4}{\nu^2}) - (1-\frac{2}{\nu})( |g^{(1)}(\vec{m})|^2+|g^{(1)}(\vec{n})|^2+|g^{(1)}(\vec{m}+\vec{n})|^2 )}{ 2\abs{g^{(1)}(\vec{m})}\abs{g^{(1)}(\vec{n})}\abs{g^{(1)}(\vec{m}+\vec{n})} } \label{eq:TriplePhase} 
 \end{multline} 
The inverse cosine of this expression is known as the \emph{closure phase}, which we represent via the symbol \begin{equation} 
    \label{eq:PhiDef} \Phi(\vec{m},\vec{n}) = \pm \left[ \phi(\vec{m}+\vec{n}) - \phi(\vec{m}) - \phi(\vec{n}) \right] 
 \end{equation} 
Just as in the Siegert Relation, the third-order correlation function encodes the phase $\phi(\vec{q})$ at pixels in $\vec{k}$-space beyond the physical spatial extent of the detector ($\abs{\vec{q}_{\text{max}}} = 2\abs{\vec{k}_{\text{max}}}$) as depicted in Fig.~\ref{fig:Intro}. Together, the double and triple correlations allow retrieval of the equivalent of a coherent diffraction pattern \emph{and} its phase across an area of $\vec{k}$-space four times larger than the area of detector coverage~\cite{Classen_2017, Trost_2020}.

\section{Phase Retrieval \label{sect:Algorithm}}
Equation~\ref{eq:PhiDef} for the \emph{closure phase} $\Phi(\vec{m},\vec{n})$ is a difference equation which can be used like a discrete derivative to estimate the slope of $\phi(\vec{m})$ between pixels separated by $\vec{n}$. The anti-symmetry of the phase pins $\phi(\vec{q}=\vec{0}) = 0$. Since overall translation in real-space results in phase ramps in reciprocal space, we can estimate the value of the phase at a nearest-neighbor pixel of $\phi(\vec{q}=\vec{0}) = 0$ without loss of generality. This estimate can be refined later by seeking to minimize the total error (see Section \ref{sect:NumericalAlgorithm}) of all pixels and treating the initial value as a parameter. The difference equation and calculated $\Phi(\vec{m},\vec{n})$ from experimental data reveals the value of the phase at the next-nearest-neighbor pixels and so forth until the phase on the entire pixel array has been calculated. 

Once the second pixel of $\phi$ (next-nearest neighbor) in any direction is calculated, the interval of the difference equation ($\vec{n}$) may be increased to find the slope between every other (instead of every) pixel in the same direction. Essentially, the phase values calculated for pixels near the origin constrain the possible phase values of pixels far from the origin.

\begin{figure}[htb]
\centering\includegraphics[width=\textwidth]{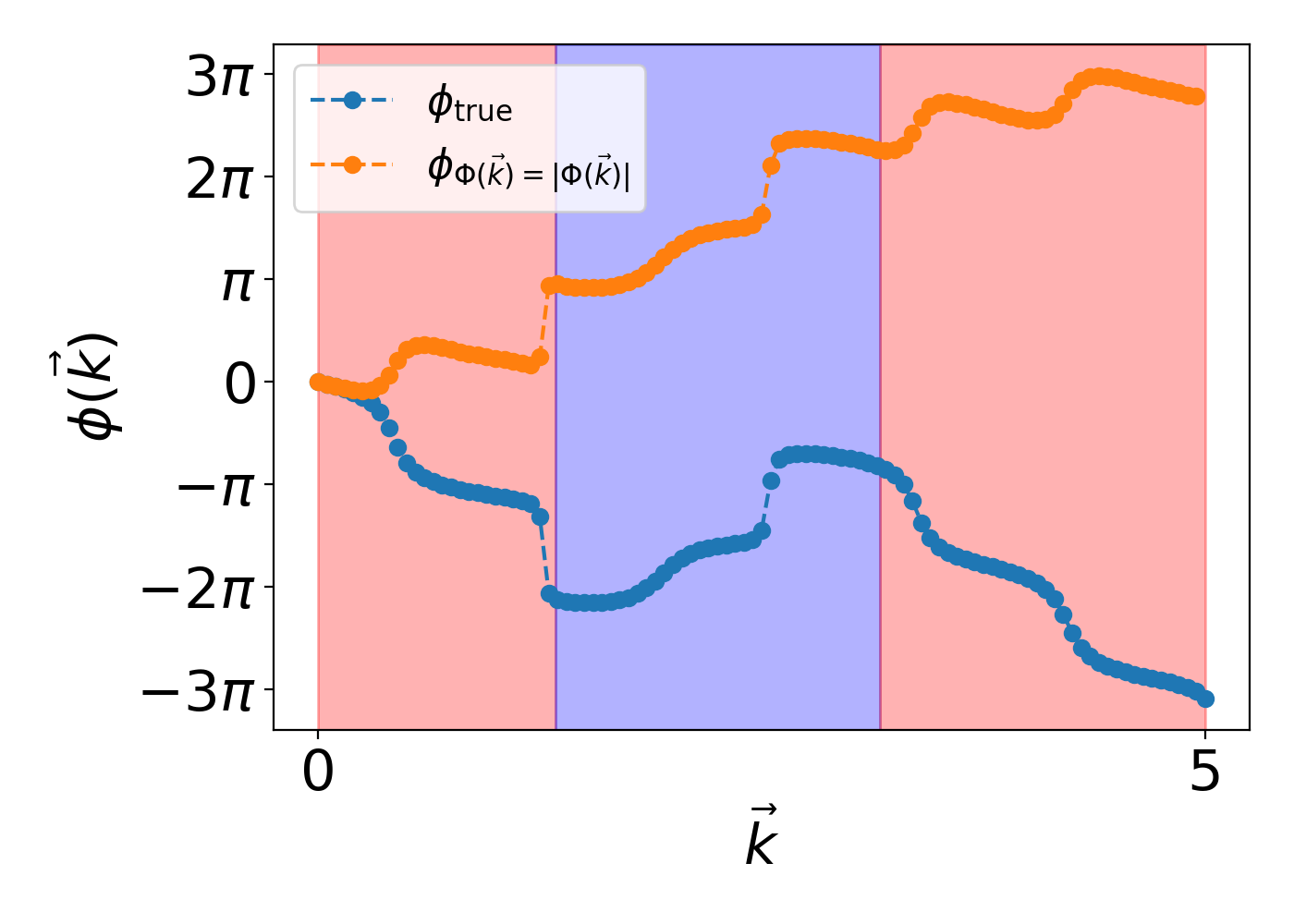}
\caption{Phase retrieved for a 1D detector assuming $\Phi(\vec{m},\vec{n}) = \abs{\Phi(\vec{m},\vec{n})}$. The correct contour is recovered with the wrong slope in the sections highlighted in red. The blue highlight section has both the correct shape and slope, but is offset due to incorporation of incorrect $\phi$ early in the solution.
\label{fig:Simple}}
\end{figure}

\subsection{Determining the sign of the closure phase $\text{sgn}(\Phi)$}
Due to the sign ambiguity of the inverse cosine, every datum from $\Phi(\vec{m},\vec{n})$ points to two possible values of the phase $\phi(\vec{m}+\vec{n})$ 
 \begin{equation} 
    \phi(\vec{m}+\vec{n}) = + \Phi(\vec{m},\vec{n}) + \phi(\vec{m}) + \phi(\vec{n}) \equiv \theta_+  
 \end{equation} 
or 
 \begin{equation}  
    \phi(\vec{m}+\vec{n}) = - \Phi(\vec{m},\vec{n}) + \phi(\vec{m}) + \phi(\vec{n}) \equiv \theta_-  
 \end{equation} 
for any $\vec{m}$ and $\vec{n}$. Assuming a global sign often leads to an incorrect slope for $\phi$, as shown in Fig.~\ref{fig:Simple}. The fact that multiple values of $\Phi(\vec{m},\vec{n})$ relate to the value of the phase at a single pixel allows us to determine the proper sign of $\Phi(\vec{m},\vec{n})$ for each $\vec{m}$ and $\vec{n}$. 

Suppose for a given pixel at $\vec{u}$ there exist $N$ sets $(\vec{m}$, $\vec{n})$ in $\Phi(\vec{m}, \vec{n})$ for which $\vec{m} + \vec{n} = \vec{u}$. Each set offers a pair of possible values for $\phi(\vec{u})$, giving $2N$ possible values for $\phi(\vec{u})$ altogether. We know that each and every one of the $N$ pairs contains the correct value, so comparing the $N$ pairs should reveal it. Ideally, the correct value is included $N$ times between the $N$ pairs and is found simply by taking the intersection of all pairs. Next, we show a simple 1D example to illustrate the principle. 

\subsubsection{Phase Retrieval 1D Example \label{sect:1D_Example}}
Suppose we have a $\phi(m) = (0,1,-3,-1,4,2,7)$ for $m=[0,6]$. We can calculate a matching $\Phi(m,n)$ to which we add a sign ambiguity. (We will render our $\Phi(m,n)$ matrix with $mn$-axes such that the origin is in the bottom left corner, i.e. so that $\Phi(1,1) = 5$). 
 \begin{equation}
    |\Phi| = \left(\begin{matrix}
    0\\
    0&4\\
    0&3&6\\
    0&4&6&9\\
    0&1&10&6&6\\
    0&5&1&4&3&4\\
    0&0&0&0&0&0&0\\
    \end{matrix}\right)
 \end{equation}
The upper right corner has no $m+n\leq6$ and here the difference equation \ref{eq:PhiDef} is undefined.

For purposes of this demonstration, let us assume we have correctly estimated $\phi(1) = 1$. Then, we can try to determine $\phi(2)$ via 
 \begin{equation} \Phi(1,1) = |\phi(2) - 2 \phi(1)| = 5 \end{equation} which gives the following options 
 \begin{equation} \phi(2) = 5 + 2\phi(1) = 7 \end{equation} or 
 \begin{equation} \phi(2) = -5 + 2\phi(1) = -3 \end{equation}
Since this is a 1D example, there is no other point $(m,n)$ such that we can deduce the correct value using $\Phi(m,n)$. The same is true for the calculation of $\phi(3)$ since we only have the information $\Phi(1,2)=1$ available to us. In a 2D phase retrieval we would have more values of $\Phi$ available to us to solve these pixels close to the origin, but for this 1D example, we will follow the branch we know to be correct with $\phi(2) = -3$ and $\phi(3) = -1$.

Now the core principle of our algorithm will be demonstrated. We  want to find $\phi(4)$ using $\phi(1)$, $\phi(2)$, $\phi(3)$, $\Phi(2,2)$, and $\Phi(1,3)$. This is the first pixel for which we have two constraints on $\phi(4)$ via $\Phi(2,2)$ and $\Phi(1,3)$ \begin{equation}  \phi(4) = \pm \Phi(2,2) +2\phi(2) = \pm 10 + -6 \end{equation}  \begin{equation} \phi(4) = \pm \Phi(1,3) + \phi(1) + \phi(3) = \pm 4 + 0  \end{equation}  which gives us four possible values of $\phi(3)$ in two pairs labeled by the points on the $\Phi$ matrix that are associated with ($\Phi(2,2) \rightarrow \{4, -16\}$ and $\Phi(1,3) \rightarrow \{4,-4\}$). If we assume that our calculation of $\phi$ up to $\phi(3)$ is accurate, then we should expect that the correct value of $\phi(n=4)$ is contained in \emph{both} pairs. Examining the intersection of $(4,-4)$ and $(4,-16)$, we may infer that $\phi(4)=4$.

Similarly, we may determine the correct value and sign of $\phi(5)$ via two constraints from $\abs{\Phi}$.
 \begin{equation} \phi(5) = \pm \Phi(2,3) + \phi(2) + \phi(3) = \pm 6 + -4 \rightarrow \{2,-10\} \end{equation}  
 \begin{equation} \phi(5) = \pm \Phi(1,4) + \phi(1) + \phi(4) = \pm 3 + 5 \rightarrow \{8,2\} \end{equation} 
Again, the intersection of the pairs of possible solutions gives the correct answer $\phi(5) = 2$. 

The last entry requires finding the intersection among three pairs since there are three non-redundant pieces of data in $\abs{\Phi}$ we may use. 
 \begin{equation} \phi(6) = \pm \Phi(3,3) + \phi(3) + \phi(3) = \pm 9 -2 \rightarrow \{7,-11\} \end{equation}  
 \begin{equation} \phi(6) = \pm \Phi(2,4) + \phi(2) + \phi(4) = \pm 6 +1 \rightarrow \{7,-5\} \end{equation}
 \begin{equation} \phi(6) = \pm \Phi(1,5) + \phi(1) + \phi(2) = \pm 4 +3 \rightarrow \{7,-1\} \end{equation}
The intersection of the three pairs leads us to conclude that $\phi(6) = 7$.

\subsection{Numerical Algorithm \label{sect:NumericalAlgorithm}}
In practice, the intersection of solution pairs is never exact and a numerical estimation subject to input noise is required. We devised an algorithm to accurately find the intersection of all pairs of possible solutions. For each $\vec{u} = \vec{m} + \vec{n}$ there are, in general, multiple sets of $(\vec{m}, \vec{n})$ positions with previously calculated phases.  Each of these sets generates a pair of solutions for the two signs of $\Phi(\vec{m}, \vec{n})$, say $\theta_{+,i}$ and $\theta_{-,i}$. As seen in the example above, for every $i$, one of these two values is approximately the same (mod $2 \pi$). 

\begin{figure}[htb]\centering\includegraphics[width=\textwidth]{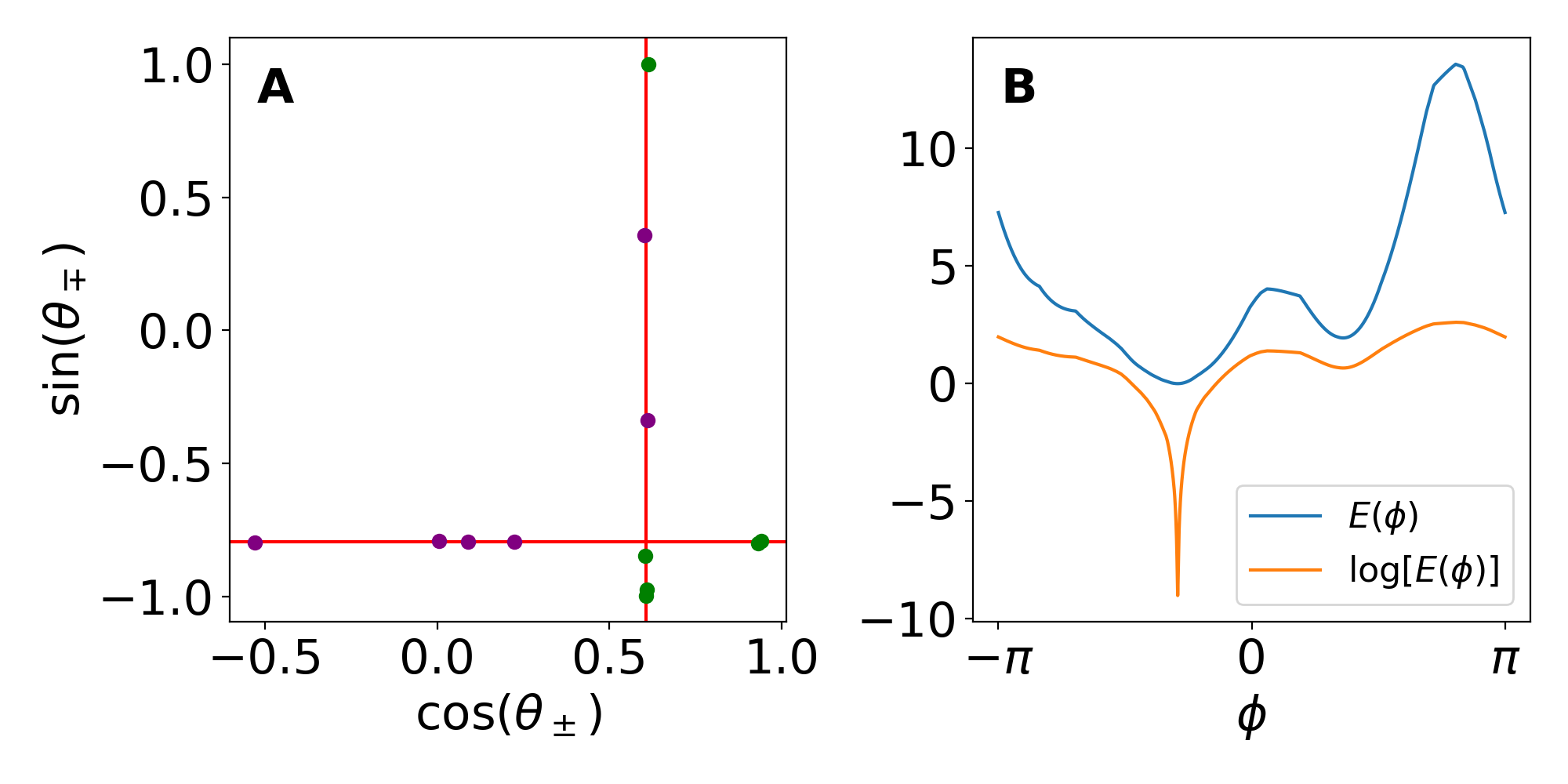}
\caption{Successful minimization of an example error function $\log E(\phi)$ in (B) finds the correct intersection of the set of $\theta_\pm$ in figure (A). Points in green are the ordered pairs $(\cos \theta_+, \sin \theta_-)$ while points in purple are the ordered pairs $(\cos \theta_-, \sin \theta_+)$, as described in the main text.
\label{fig:PhaseFitting}}
\end{figure}

A visually instructive way to determine this common value is to plot each of these pairs as points in a two-dimensional plane, both as  $(\theta_+, \theta_-)$ and $(\theta_-, \theta_+)$. Thus, either the horizontal and vertical components of each point are common with all the others. Graphically, this means that the points (approximately) form a vertical and horizontal line intersecting at the true solution $(\phi(\vec{u}), \phi(\vec{u}))$. In order to stay in a bounded domain, it is simpler to consider ordered pairs $(\cos(\theta_+), \sin(\theta_-))$ and $(\cos(\theta_-), \sin(\theta_+))$ to constrain the search to $\phi \in [-\pi,\pi]$ and remove $2\pi$ offsets of the value of $\theta_\pm$ (see Fig.~\ref{fig:PhaseFitting}A). Finding the intersection given some noise in the data $\theta_\pm$ then amounts to the minimization of the error function

 \begin{equation} 
    E(\phi) = \sum_i \min \left[ (\cos(\theta_\pm)_i - \cos(\phi))^2, (\sin(\theta_\mp)_i - \sin(\phi))^2 \right] \label{eq:ErrorFunc}
 \end{equation} 
where the sum is over the $N$ pairs of possible solutions for our chosen $\vec{u}$. The optimal value is the desired value of $\phi$ at $\vec{u}$, $\phi(\vec{u}) = \phi_{\text{opt}}$.

The landscape of the error function $E(\phi)$ presents challenges for conjugate gradient optimization because it contains multiple local minima separated by large barriers. An example for a random pixel in a 2D phase array is shown in Fig.~\ref{fig:PhaseFitting}B. Since the value of the phase needs to be optimized for each pixel on the detector, a rapid and accurate method of determining the absolute minimum is desired. This is most straightforwardly accomplished by supplying an optimization algorithm with the minimum value of $\log E(\phi)$ on a grid in $[-\pi, \pi]$; the logarithm increases the contrast of the absolute minima significantly, allowing accurate calculation of an initial guess for the conjugate gradient optimizer. The optimizer polishes the brute-force search to a precise final value.

\begin{figure*}[htb]
\centering\includegraphics[width=\textwidth]{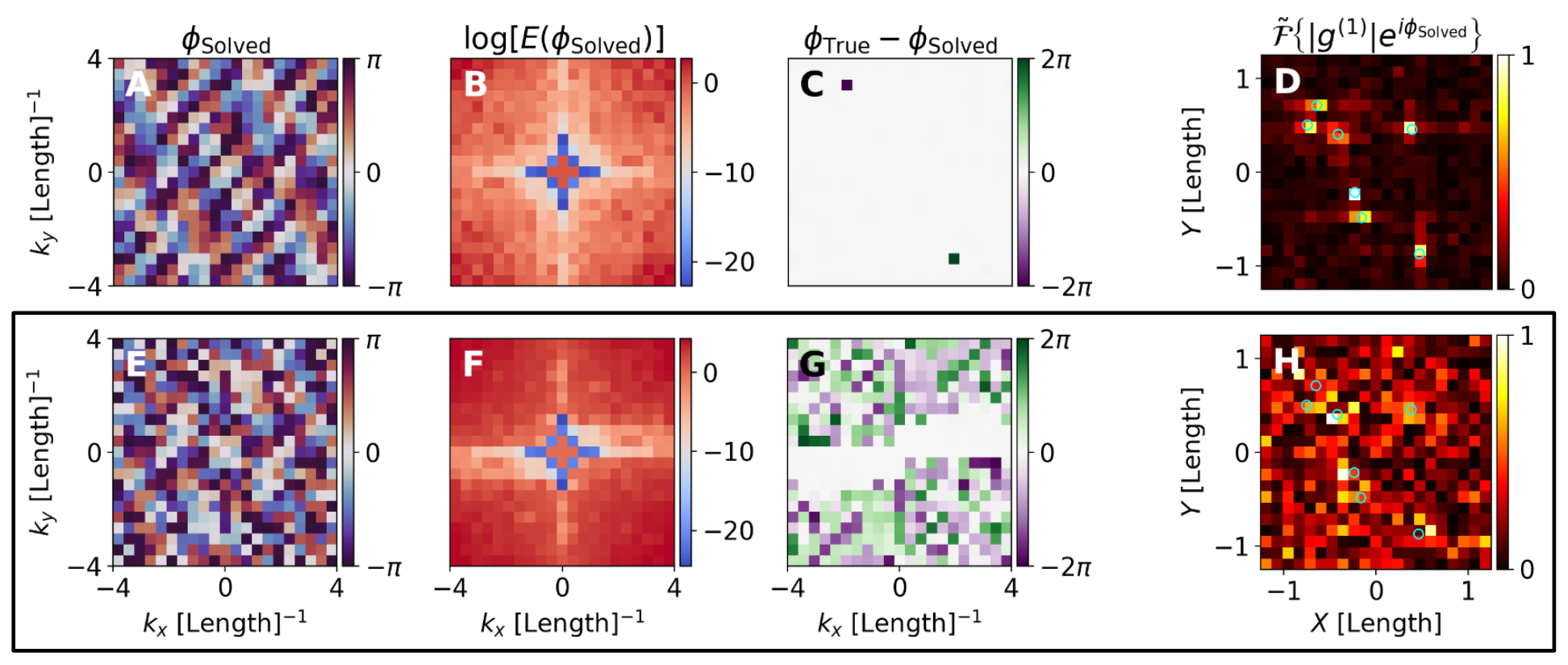}
\caption{Exact ab initio phase retrieval from triple correlations both with (A-D) and without (boxed, E-H) alternate value toggling. (A) shows the phase retrieved from the third-order correlations used to produce the object image via Fourier inversion in (D). The true positions of the atoms used in the simulation are indicated by light blue circles in (D) and (H). (B) shows the error values calculated for each pixel during phase retrieval and (C) shows the difference between the true and retrieved phase values. (E-H) shows the same plots for the same set of atoms and number of shots but without alternate phase value toggling. When alternate phase values at pixels adjacent to spikes in the error function are toggled correctly as in (B), the error is reduced and the difference between the retrieved and true values is small. When alternates are not toggled correctly as in (F), the error is large and faithful structure retrieval is less likely. Phase information past the physical edge of the detector ($\abs{\vec{k}_{\text{max}}} = 2$ in this example) is retrieved via the triple correlations, enhancing real space resolution that would be measured in a typical diffraction experiment.
\label{fig:2DPhaseRetrieval}}
\end{figure*}

Since the value of $\phi(\vec{m}+\vec{n})$ depends on previous values of $\phi(\vec{m})$ and $\phi(\vec{n})$, it is especially important that values calculated early in the retrieval are accurate. Depending on the quality of the data in $\Phi(\vec{m},\vec{n})$, the error function may present multiple deep local minima which cause the algorithm to, initially, choose an incorrect value of $\phi$. In this case, $\log[E(\phi)]$ for subsequent pixels in the retrieval sharply increases, indicating that at least one previous pixel has $\phi$ assigned incorrectly. Plotting the total fit error for all pixels indicates the location of problematic pixels which require resolving by toggling candidate phase values until the total error of all pixels is minimized. The consequences of toggling alternate values of $\phi$ near the origin to minimize the error across all pixels are illustrated nicely in Fig.~\ref{fig:2DPhaseRetrieval} by comparing the boxed and unboxed figures.

\section{Results}

Using Equation \ref{eq:TriplePhase} and the algorithm described in Section \ref{sect:Algorithm}, we can calculate the Fourier phase from triple correlation data and compare the result to the true value. Fig.~\ref{fig:2DPhaseRetrieval} shows the results of phase retrieval in a simulation with a 2D pixel detector. Note that in regions where the error Fig.~\ref{fig:2DPhaseRetrieval}B is small, the solved phase matches the true phase quite well. In this example, sufficient phase information is retrieved to fully resolve the seven simulated atoms (blue circles in Fig.~\ref{fig:2DPhaseRetrieval}D) with only $10^4$ shots. The Fourier inversion was performed using the phase retrieved via our algorithm from the third-order correlation function and $\abs{g^{(1)}}$ calculated via the second-order correlation function. No use of coherent diffraction data was required. 

Acquiring additional shots significantly improves the fidelity of phase retrieval. The primary practical limit on phase retrieval via the triple correlations is the computation of the bispectrum which, for a 2D detector, requires storage of a $(N_{\text{pix}} \times N_{\text{pix}})^2$ floating point array. For large detectors, the bispectrum can rapidly consume all available memory on small workstations. The $11\times11$ detector simulated in Fig.~\ref{fig:2DPhaseRetrieval} with $10^4$ shots was chosen as a reasonable compromise between memory usage, execution time, and visual impact for this demonstration on a 2015 MacBook Pro running a 2.8GHz Intel i7 processor.

\section{Conclusions}
We have described a mathematical solution to the sign problem in phase retrieval from triple correlations of fluorescent or thermal light. We provided a numerical demonstration of our method with simulated data which accurately retrieved the reciprocal space Fourier phase of an atom array. We envision this method as another potential solution to the phase problem in crystallography via photon correlations of fluorescence radiation from atoms pumped by x-ray free-electron lasers. The method presented in this paper may also prove interesting for pulse metrology~\cite{Feurer_1998, Liu_2002, Inoue_2019, Nakamura_2020}, observing many-body correlations in ultracold atomic gases~\cite{Folling_2005, Jeltes_2007, Andrews_1997}, imaging in turbid media~\cite{Bertolotti_2012, Katz_2014, Stern_2019}, and for imaging with radio telescope arrays~\cite{Wentz_2014,Dravins_2015,Nunez_2015}.

\begin{backmatter}
\bmsection{Funding}
NP acknowledges financial support from the MIT International Science and Technology Initiatives (MISTI) program and the Hertz Foundation. HNC acknowledges support from DESY (Hamburg, Germany), a member of the Helmholtz Association HGF. This work is supported by the Cluster of Excellence 'CUI: Advanced Imaging of Matter' of the Deutsche Forschungsgemeinschaft (DFG) - EXC 2056 - project ID 390715994. This work was funded by Deutsche Forschungsgemeinschaft (DFG, German Research Foundation) - 491245950.

\bmsection{Acknowledgments}
NP thanks Anlong Chua for helpful discussions regarding Eq. \ref{eq:ErrorFunc} and Fabian Trost for helpful discussions regarding computation of the bispectrum. 

\bmsection{Disclosures}
The authors declare no conflicts of interest.

\bmsection{Data Availability Statement}
The data and source code which support the results in this paper are available on Github at \url{https://github.com/npeard/CorrSpeck}.

\bmsection{Supplemental document}
See Supplement 1 for supporting content. 

\end{backmatter}

\bibliography{biblio}

\clearpage

\renewcommand{\theequation}{S\arabic{equation}}
\renewcommand{\thefigure}{S\arabic{figure}}
\renewcommand{\thesection}{S\arabic{section}}
\setcounter{figure}{0}
\setcounter{section}{0}
\setcounter{equation}{0}

\section{Supplement}

\subsection{Importance of the Fourier Phase}
\begin{figure}[htb]
\centering\includegraphics[width=\textwidth]{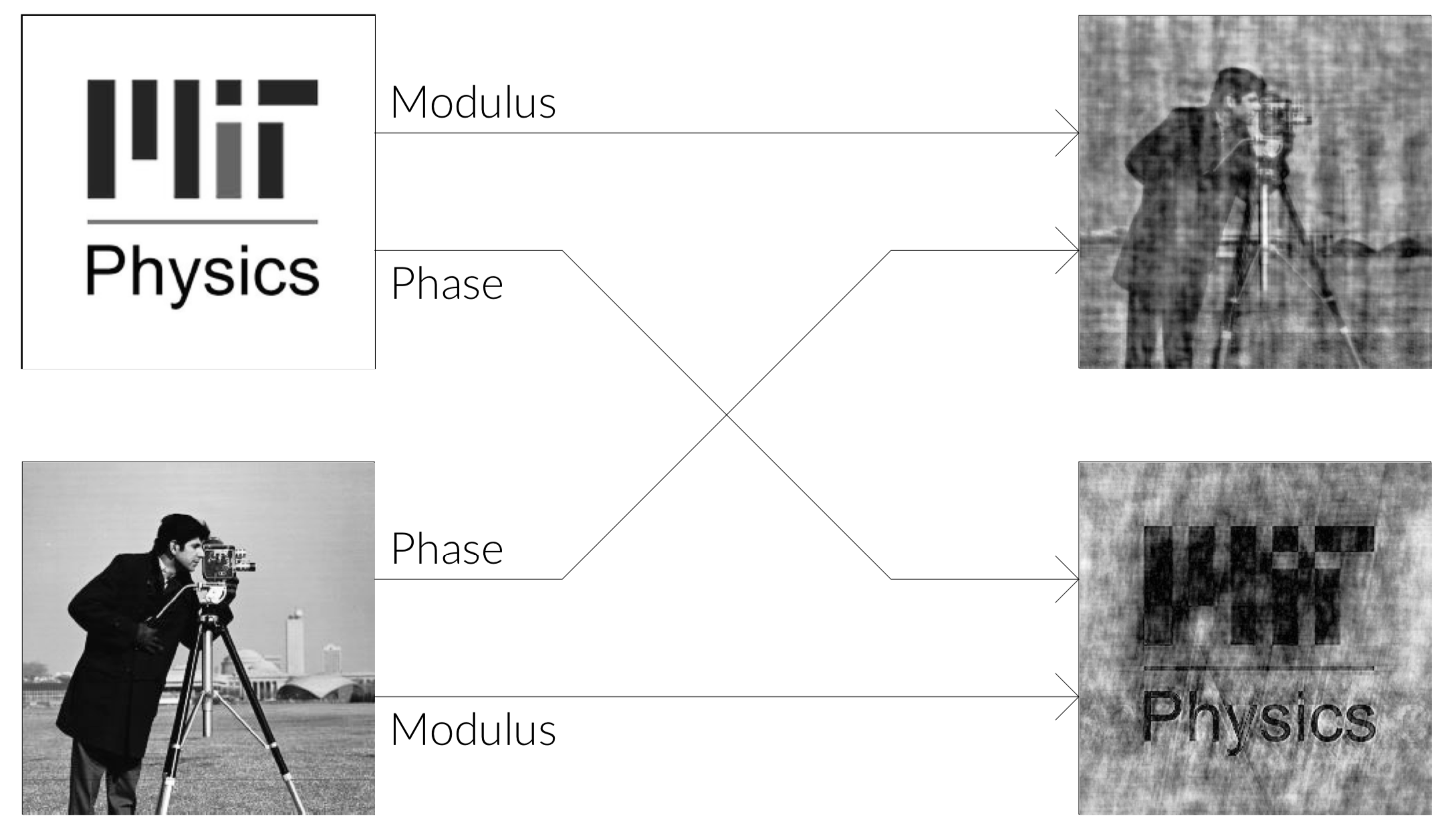}
\caption{Two images (left) are Fourier transformed. Swapping the phases of the transformed images and then executing the inverse Fourier transform (right) demonstrates the importance of the phase information in diffractive imaging. Square-law detectors (most photodetectors) cannot access the phase of the light field directly, so creative indirect techniques are required to recover the total light field.
\label{fig:PhaseImportant}}
\end{figure}

\subsection{Detection of a Photon}
The reader may wish to refer to \cite{Glauber_1963a, Glauber_1963b, Glauber_1963c, Perina_1991} in the next few sections where we derive the expression for intensity triple correlations used in the main text. 

Consider some initial state of the photon field $\ket{i}$ with some detection device in the ground state. $\ket{f}$ is the final state of the field after our detection system has been excited by a photon. The transition amplitude for this event is then 
 \begin{equation}
     A = \bra{f} \hat{\psi} \ket{i}
 \end{equation}
and the transition probability is 
 \begin{equation}
     P = \abs{A}^2 = \abs{ \bra{f} \hat{\psi} \ket{i} }^2
 \end{equation}
where $\psi$ is the photon field annihilation operator. The final state of the field is usually not observed directly, so we trace over the possible $\ket{f}$
 \begin{equation}
     P = \sum_f \bra{i} \hat{\psi}^\dag \ket{f}\bra{f} \hat{\psi} \ket{i} = \bra{i} \hat{\psi}^\dag \hat{\psi} \ket{i}
 \end{equation}
This short calculation shows how photon detection is analogous to a measurement of the degree of first-order coherence.
 
\subsection{First-Order Coherence in Diffraction}
In coherent diffraction experiments, we measure the degree of first-order coherence as a function of reciprocal space coordinates. 
 \begin{equation}
      G^{(1)}(\vec{k}, \vec{k}_0) = \langle \hat{\Psi}^\dag(\vec{k}) \hat{\Psi}(\vec{k}_0) \rangle
 \end{equation}
where $\hat{\Psi}(\vec{k}) = \sum_i^\nu e^{-\iu \vec{k} \cdot \vec{r}_i} \hat{a}_i$ is the photon field operator summing over the number of scatterers ($\nu$). 
 \begin{equation}
      G^{(1)}(\vec{k}, \vec{k}_0) = \sum_{ij}^\nu e^{\iu \vec{k} \cdot \vec{r}_j} e^{-\iu \vec{k}_0 \cdot \vec{r}_i} \langle \hat{a}_j^\dag \hat{a}_i \rangle \delta_{ij}
 \end{equation}
In this picture of elastic scattering, a photon in mode $\vec{k}_0$ is destroyed at site $\vec{r}_i$ and a new photon of the same energy is created in the mode $\vec{k}$ at site $\vec{r}_j = \vec{r}_i$. If we define the momentum transfer vector $\vec{q} = \vec{k} - \vec{k}_0$, then
 \begin{equation}
     G^{(1)}(\vec{q}) = G^{(1)}(\vec{k}, \vec{k}_0) = \sum_{i}^\nu e^{\iu \vec{q} \cdot \vec{r}_i} \langle \hat{a}_i^\dag \hat{a}_i \rangle 
 \end{equation}
For elastic scattering, we can generally regard the state of the light field to be classical. Furthermore, we consider a system with only one atomic species such that $\alpha_i = \alpha_j = \alpha$, and
 \begin{equation}
     G^{(1)}(\vec{q}) = \abs{\alpha}^2 \sum_{i}^\nu e^{\iu \vec{q} \cdot \vec{r}_i} \label{app:eq:G1}
 \end{equation}
It follows that $G(\vec{0}) = \abs{\alpha}^2 \nu$ so the normalized correlation function is
 \begin{equation}
     g^{(1)}(\vec{q}) = \frac{G^{(1)}(\vec{q})}{G^{(1)}(\vec{0})} = \frac{1}{\nu}  \sum_k^\nu e^{\iu \vec{q} \cdot \vec{r}_k}
 \end{equation}
Photodetectors measure only the fringe visibility (the light intensity), eliminating the phase information of the first-order coherence 
 \begin{equation}
    \abs{g^{(1)}(\vec{q})} = \frac{1}{\nu} \abs{ \sum_k^\nu e^{\iu \vec{q} \cdot \vec{r}_k} } \label{app:eq:g1ModSquare}
 \end{equation}

\subsection{Pair Correlation Function \label{app:PairCorrTheory}}
Light with no first-order coherence (e.g., fluorescence) can still exhibit second-order coherence. The pair-wise correlation function is written in second-quantization as
\begin{equation}
    G^{(2)}(\vec{k}_1,\vec{k}_2) = \langle \hat{\psi}^\dag(\vec{k}_1) \hat{\psi}^\dag(\vec{k}_2) \hat{\psi}(\vec{k}_1) \hat{\psi}(\vec{k}_2) \rangle
\end{equation} where the angle brackets indicate the expectation value of a quantum state.
Expanding single-particle states in a single mode of momentum-space
\begin{equation}
    \hat{\psi}(\vec{k}) = \sum_i^\nu e^{-\iu \vec{k} \cdot \vec{r}_i} e^{-i \phi_i} \hat{a}_i = \sum_i^\nu \mathcal{E}_i(\vec{k}) \hat{a}_i
\end{equation}
with $\phi_i \in [0,2\pi)$ a random phase that varies slower than the coherence time. In this picture, each atom at $\vec{r}_i$ emits a photon into the mode $\vec{k}$ with phase $\phi_i$. 

The pair-wise correlation averaged over an ensemble of measurements (shots) is 
\begin{equation}
    \left\{ G^{(2)}(\vec{k}_1,\vec{k}_2) \right\} =  \left\{ \sum_{ijkl}^\nu \mathcal{E}_i^*(\vec{k}_1) \mathcal{E}_j^*(\vec{k}_2) \mathcal{E}_k(\vec{k}_1) \mathcal{E}_l(\vec{k}_2) \langle \hat{a}_i^\dag \hat{a}_j^\dag \hat{a}_k \hat{a}_l \rangle \right\}
\end{equation} 
The outer braces notate the ensemble mean.

There are two cases where the ensemble mean pair-wise correlation is non-zero:
 \begin{equation}
    i=k, \quad j=l \label{app:eq:Case1}
 \end{equation}
 \begin{equation}
    i=l, \quad j=k \label{app:eq:Case2}
 \end{equation} 
We contract the fields accordingly and rewrite the correlation function as
 \begin{multline}
    \left\{ G^{(2)}(\vec{k}_1,\vec{k}_2) \right\} =  \\ \Bigg\{ \sum_{ij}^\nu \mathcal{E}_i^*(\vec{k}_1) \mathcal{E}_j^*(\vec{k}_2) \mathcal{E}_i(\vec{k}_1) \mathcal{E}_j(\vec{k}_2) \langle \hat{a}_i^\dag \hat{a}_j^\dag \hat{a}_i \hat{a}_j \rangle \\ + \mathcal{E}_i^*(\vec{k}_1) \mathcal{E}_j^*(\vec{k}_2) \mathcal{E}_j(\vec{k}_1) \mathcal{E}_i(\vec{k}_2) \langle \hat{a}_i^\dag \hat{a}_j^\dag \hat{a}_j \hat{a}_i \rangle \\ - \sum_i^\nu \mathcal{E}_i^*(\vec{k}_1) \mathcal{E}_i^*(\vec{k}_2) \mathcal{E}_i(\vec{k}_1) \mathcal{E}_i(\vec{k}_2) \langle \hat{a}_i^\dag \hat{a}_i^\dag \hat{a}_i \hat{a}_i \rangle \Bigg\} \label{app:eq:G2Cases}
 \end{multline}
By writing out the first two terms with $i,j$ indices over the number of emitters we double count the case that $i=j$. In Equation \ref{app:eq:G2Cases}, we have subtracted this case once in the last term. 

The quantum-classical correspondence between field annihilation/creation operators and classical waves ($\hat{A}^{(+)} \leftrightarrow V$ and $\hat{A}^{(-)} \leftrightarrow V^*$) with normal ordered correlation functions is implemented with the use of the coherent states $\ket{\{\alpha\}}$ \cite{Perina_1991}. Fock states have no corresponding classical state and must be treated separately.

\subsubsection{Classical Light}
Consider the case that each atom emits a single coherent field so that the initial state of the field is 
 \begin{equation}
     \ket{\{\alpha\}} = \prod_\lambda^\nu \ket{\alpha_\lambda}
 \end{equation}
 \begin{equation}
     \hat{a}_\sigma \ket{\{\alpha\}} = \alpha_\sigma \ket{\{\alpha\}}
 \end{equation}
 where $\ket{\alpha_\lambda}$ is the state contributed by the $\lambda$th atom.
Note that
 \begin{equation}
     \bra{\{\alpha\}} \hat{a}_\mu^\dag \hat{a}_\sigma \ket{\{\alpha\}} = \alpha_\mu^* \alpha_\sigma
 \end{equation}
since the coherent states are eigenstates of $\hat{a}$. Because the correlation function operator is normal ordered, we can directly resume from Equation \ref{app:eq:G2Cases} with
 \begin{multline}
     \left\{ G^{(2)}(\vec{k}_1,\vec{k}_2) \right\} =  \Bigg\{ \sum_{ij}^\nu \mathcal{E}_i^*(\vec{k}_1) \mathcal{E}_j^*(\vec{k}_2) \mathcal{E}_i(\vec{k}_1) \mathcal{E}_j(\vec{k}_2) \abs{\alpha_i}^2 \abs{\alpha_j}^2 \\ + \mathcal{E}_i^*(\vec{k}_1) \mathcal{E}_j^*(\vec{k}_2) \mathcal{E}_j(\vec{k}_1) \mathcal{E}_i(\vec{k}_2) \abs{\alpha_i}^2 \abs{\alpha_j}^2  - \sum_i^\nu \mathcal{E}_i^*(\vec{k}_1) \mathcal{E}_i^*(\vec{k}_2) \mathcal{E}_i(\vec{k}_1) \mathcal{E}_i(\vec{k}_2) \abs{\alpha_i}^4 \Bigg\}
 \end{multline}
Now, the ensemble cancellation of random phases in the $\mathcal{E}_i(k)$ gives the following approximation
 \begin{multline}
     N \left\{ G^{(2)}(\vec{k}_1,\vec{k}_2) \right\} \approx  \sum_{ij}^\nu \abs{\mathcal{E}_i(\vec{k}_1)}^2 \abs{\mathcal{E}_j(\vec{k}_2)}^2 \abs{\alpha_i}^2 \abs{\alpha_j}^2 \\ + \mathcal{E}_i^*(\vec{k}_1) \mathcal{E}_j^*(\vec{k}_2) \mathcal{E}_j(\vec{k}_1) \mathcal{E}_i(\vec{k}_2) \abs{\alpha_i}^2 \abs{\alpha_j}^2 - \sum_i^\nu \abs{\mathcal{E}_i(\vec{k}_1)}^4 \abs{\alpha_i}^4 
 \end{multline}
Supposing identical emission from each atom, $\alpha_i = \alpha_j = \alpha$, and recalling $\abs{\mathcal{E}_i(k)}^2 = 1$, we can reduce the sums over constant terms 
 \begin{equation}
     N \left\{ G^{(2)}(\vec{k}_1,\vec{k}_2) \right\} \approx  \nu^2 \abs{\alpha}^4 - \nu \abs{\alpha}^4  + \sum_{ij}^\nu \mathcal{E}_i^*(\vec{k}_1) \mathcal{E}_j^*(\vec{k}_2) \mathcal{E}_j(\vec{k}_1) \mathcal{E}_i(\vec{k}_2) \abs{\alpha}^4 
 \end{equation}
We can rewrite the last term
 \begin{equation}
     \sum_{ij}^\nu \mathcal{E}_i^*(\vec{k}_1) \mathcal{E}_i(\vec{k}_2) \mathcal{E}_j^*(\vec{k}_2) \mathcal{E}_j(\vec{k}_1) = 
     \sum_{ij}^\nu e^{\iu (\vec{k}_1-\vec{k}_2)\cdot \vec{r}_j} e^{-\iu (\vec{k}_1-\vec{k}_2) \cdot \vec{r}_i} 
     = \abs{\sum_j^\nu e^{\iu \vec{q} \cdot \vec{r}_j}}^2
 \end{equation}
where $\vec{q} = \vec{k}_1-\vec{k}_2$. Normalizing the entire correlation function by dividing out the square of
 \begin{equation}
     N \left\{ \langle \hat{\psi}^\dag(\vec{k}_0) \hat{\psi}(\vec{k}_0) \rangle \right\} = N \left\{ \sum_{ij}^\nu e^{\iu \vec{k}_0 \cdot (\vec{r}_i - \vec{r}_j)} e^{\iu (\phi_i - \phi_j)} \langle \hat{a}_i^\dag \hat{a}_j \rangle \right\} \\ \approx \sum_{ij}^\nu \langle \hat{a}_i^\dag \hat{a}_j \rangle \delta_{ij} = \sum_i^\nu \langle \hat{a}_i^\dag \hat{a}_i \rangle = \nu \abs{\alpha}^2
 \end{equation}
we can express the normalized second-order correlation function
 \begin{equation}
     g^{(2)}(\vec{k}_1,\vec{k}_2) = \frac{ \left\{ G^{(2)}(\vec{k}_1,\vec{k}_2) \right\} }{ \left\{ \langle \hat{\psi}^\dag(\vec{k}_0) \hat{\psi}(\vec{k}_0) \rangle \right\}^2 } 
 \end{equation}
as 
 \begin{equation} 
     g^{(2)}(\vec{k}_1,\vec{k}_2) \approx \\ 1 - \frac{1}{\nu} + \frac{1}{\nu^2} \abs{\sum_j^\nu e^{\iu \vec{q} \cdot \vec{r}_j} }^2 
 \end{equation}
Observe that the last term is identical to the modulus square first-order correlation function from Equation \ref{app:eq:g1ModSquare}. Apparently, the coherent diffraction pattern is recovered from photon pair correlations of fluorescence speckle intensities.
 \begin{equation} 
     g^{(2)}(\vec{k}_1,\vec{k}_2) \approx \\ 1 - \frac{1}{\nu} + \abs{ g^{(1)}(\vec{q}) }^2
 \end{equation}

The imperfect cancellation of random phases causes the "phase noise" of Hanbury Brown and Twiss. The more atoms included, the longer random walk in the complex plane requires additional shots to reach the same level of phase noise as with fewer atoms.

\subsubsection{Quantum Light}

Consider the case that each atom emits a single photon so that the initial state of the field is defined
 \begin{equation}
     \ket{\{n\}} = \prod_\lambda^\nu \ket{n_\lambda}
 \end{equation}
 \begin{equation}
     \hat{a}_\sigma \ket{\{n\}} = \sqrt{n_\sigma} \ket{\{n_1,n_2,...,n_\sigma - 1, n_{\sigma+1}, ...\}}
 \end{equation}
where $\ket{n_\lambda}$ is the state contributed by the $\lambda$th atom. Note that 
 \begin{equation}
     \bra{\{n\}} \hat{a}_\mu^\dag \hat{a}_\sigma \ket{\{n\}} = \sqrt{n_\mu} \sqrt{n_\sigma} \delta_{\mu \sigma}
 \end{equation}
requires an exact match of creation and annihilation operator indices to be non-zero, since Fock states form an orthonormal basis and are not eigenstates of $\hat{a}$ or $\hat{a}^\dag$.

For coherent states, the normal-ordered correlation operator is already diagonalized. For Fock states, this is not so. The commutation relation $[\hat{a}_i, \hat{a}_j^\dag] = \delta_{ij}$ allows us to diagonalize the cases \ref{app:eq:Case1} and \ref{app:eq:Case2}
 \begin{equation}
    \langle \hat{a}_i^\dag \hat{a}_j^\dag \hat{a}_i \hat{a}_j \rangle = \langle \hat{a}_i^\dag \hat{a}_i \hat{a}_j^\dag \hat{a}_j \rangle - \langle \hat{a}_i^\dag \hat{a}_j \delta_{ij} \rangle
 \end{equation}
 \begin{equation}
    \langle \hat{a}_i^\dag \hat{a}_j^\dag \hat{a}_j \hat{a}_i \rangle = \langle \hat{a}_j^\dag \hat{a}_j \hat{a}_i^\dag \hat{a}_i \rangle - \langle \hat{a}_j^\dag \hat{a}_i \delta_{ij} \rangle
 \end{equation}
as well as the double-counting correction term in \ref{app:eq:G2Cases}
 \begin{equation}
     \langle \hat{a}_i^\dag \hat{a}_i^\dag \hat{a}_i \hat{a}_i \rangle = \langle \hat{a}_i^\dag \hat{a}_i \hat{a}_i^\dag \hat{a}_i \rangle - \langle \hat{a}_i^\dag \hat{a}_i \rangle
 \end{equation}
Writing the expectation value $\langle \hat{a}_i^\dag \hat{a}_i \rangle = n_i$ as the occupation number of the mode emitted by the $i$th atom, the diagonalized second order correlation function (Eq. \ref{app:eq:G2Cases}) reads
 \begin{multline}
    \left\{ G^{(2)}(\vec{k}_1,\vec{k}_2) \right\} =  \Bigg\{ \sum_{ij}^\nu  n_i   n_j   \abs{\mathcal{E}_i(\vec{k}_1)}^2 \abs{\mathcal{E}_j(\vec{k}_2)}^2 + \\ n_i   n_j  \mathcal{E}_i^*(\vec{k}_1) \mathcal{E}_i(\vec{k}_2) \mathcal{E}_j^*(\vec{k}_2) \mathcal{E}_j(\vec{k}_1)  - \sum_i^\nu  n_i^2  \abs{\mathcal{E}_i(\vec{k}_1)}^4 - \sum_i^\nu  n_i  \abs{\mathcal{E}_i(\vec{k}_1)}^4 \Bigg\}  
 \end{multline} 
Supposing single photon emission from identical atoms, $n_i = n_j = 1$ and writing $\abs{\mathcal{E}_j(k_2)}^2 = 1$, we get
 \begin{equation}
    \left\{ G^{(2)}(\vec{k}_1,\vec{k}_2) \right\} =  \Bigg\{ \nu^2 - 2\nu + \sum_{ij}^\nu \mathcal{E}_i^*(\vec{k}_1) \mathcal{E}_i(\vec{k}_2) \mathcal{E}_j^*(\vec{k}_2) \mathcal{E}_j(\vec{k}_1) \Bigg\}
 \end{equation}
Normalizing by $\left\{ \hat{\psi}^\dag(\vec{k}_0) \hat{\psi}(\vec{k}_0) \right\}^2 = \nu^2$ as in the previous section, we obtain 
 \begin{equation} 
     g^{(2)}(\vec{k}_1,\vec{k}_2) = \\ 1 - \frac{2}{\nu} + \abs{ g^{(1)}(\vec{q}) }^2
 \end{equation}
which is quite similar to the case where we assume classical light. Note here, however, that the expression is an exact equality - this is due to the orthogonality of Fock states contracting only those random phases which cancel perfectly. There are no $\left\{ e^{\iu \phi_i} e^{-\iu \phi_j} \right\}$ terms where $i\neq j$, so the ensemble average over shots is not subject to the Hanbury Brown and Twiss phase noise.

\subsection{Triple Correlation Function \label{app:TripleCorrTheory}}
The correlation function between triples of photons is written in second-quantization as
 \begin{equation}
    G^{(3)}(\vec{k}_1,\vec{k}_2, \vec{k}_3) = \langle \hat{\psi}^\dag(\vec{k}_1) \hat{\psi}^\dag(\vec{k}_2) \hat{\psi}^\dag(\vec{k}_3) \hat{\psi}(\vec{k}_1) \hat{\psi}(\vec{k}_2) \hat{\psi}(\vec{k}_3) \rangle
 \end{equation}
where the field operators are defined as before.
The triple correlation averaged over an ensemble of measurements is 
 \begin{equation}
    \left\{ G^{(3)}(\vec{k}_1,\vec{k}_2, \vec{k}_3) \right\} =   \left\{ \sum_{ijklmn}^\nu \mathcal{E}_i^*(\vec{k}_1) \mathcal{E}_j^*(\vec{k}_2) \mathcal{E}_k^*(\vec{k}_3) \mathcal{E}_l(\vec{k}_1) \mathcal{E}_m(\vec{k}_2) \mathcal{E}_n(\vec{k}_3) \langle \hat{a}_i^\dag \hat{a}_j^\dag \hat{a}_k^\dag \hat{a}_l \hat{a}_m \hat{a}_n \rangle \right\} \label{app:eq:G3General}
 \end{equation} 
and again we must consider the field contractions for which the expression is non-zero. We find that there are six
 \begin{equation}
    i=l, \quad j=m, \quad k=n \label{app:eq:G3Case1}
 \end{equation}
 \begin{equation}
    i=n, \quad j=l, \quad k=m \label{app:eq:G3Case2}
 \end{equation} 
 \begin{equation}
    i=m, \quad j=n, \quad k=l \label{app:eq:G3Case3}
 \end{equation} 
 \begin{equation}
    i=l, \quad j=n, \quad k=m \label{app:eq:G3Case4}
 \end{equation} 
 \begin{equation}
    i=n, \quad j=m, \quad k=l \label{app:eq:G3Case5}
 \end{equation} 
 \begin{equation}
    i=m, \quad j=l, \quad k=n \label{app:eq:G3Case6}
 \end{equation} 

The first case (\ref{app:eq:G3Case1}) contracts fields with the same $\vec{k}$, and will thus be a constant after summation and averaging. For the remaining cases, we must carefully account for double- and triple-counting. We can quickly see that cases \ref{app:eq:G3Case4}, \ref{app:eq:G3Case5}, and \ref{app:eq:G3Case6} each contract one set of fields with the same $\vec{k}$, leaving a pair of free indices which do not contract to a constant. Each of these pairs over count \ref{app:eq:G3Case1} when the indices are equal, so we must subtract this case from each of \ref{app:eq:G3Case4}, \ref{app:eq:G3Case5}, and \ref{app:eq:G3Case6}. The cases \ref{app:eq:G3Case2} and \ref{app:eq:G3Case3}, however, have three free indices which do not produce field contractions to a constant. Firstly, we can contract any pair of the three indices in each of \ref{app:eq:G3Case2} and \ref{app:eq:G3Case3} to get the result of any one of \ref{app:eq:G3Case4}, \ref{app:eq:G3Case5}, and \ref{app:eq:G3Case6} again - this over counts each of \ref{app:eq:G3Case4}, \ref{app:eq:G3Case5}, and \ref{app:eq:G3Case6} once for each of \ref{app:eq:G3Case2} and \ref{app:eq:G3Case3}. Additionally, this overcounting correction must itself be corrected just as \ref{app:eq:G3Case4}, \ref{app:eq:G3Case5}, and \ref{app:eq:G3Case6} were above. Finally, \ref{app:eq:G3Case2} and \ref{app:eq:G3Case3} each overcount the case where all three indices are equal. All cases (\ref{app:eq:G3Case1} through \ref{app:eq:G3Case6}) over-counting corrections can be expressed as 
 \begin{gather}
     \delta_{il}\delta_{jm}\delta_{kn} \nonumber \\+ \delta_{in}\delta_{jl}\delta_{km} \left(1-\delta_{ij}\right) \left(1-\delta_{jk}\right) \left(1-\delta_{ki}\right) \nonumber \\ + \delta_{im}\delta_{jn}\delta_{kl}\left(1-\delta_{ij}\right) \left(1-\delta_{jk}\right) \left(1-\delta_{ki}\right) \nonumber \\+
     \delta_{il}\delta_{jn}\delta_{km}\left(1-\delta_{jk}\right) \nonumber \\+
     \delta_{in}\delta_{jm}\delta_{kl}\left(1-\delta_{ki}\right) \nonumber \\ +
     \delta_{im}\delta_{jl}\delta_{kn}\left(1-\delta_{ij}\right) \label{app:eq:G3Tensor}  \\ = \delta_{il}\delta_{jm}\delta_{kn} 
      \nonumber \\ + \delta_{in}\delta_{jl}\delta_{km}(1-\delta_{ij}-\delta_{jk}-\delta_{ki} + \delta_{ij}\delta_{jk} + \delta_{jk}\delta_{ki} + \delta_{ki}\delta_{ij} - \delta_{ij}\delta_{jk}\delta_{ki}) \nonumber \\ + \delta_{im}\delta_{jn}\delta_{kl} (1-\delta_{ij}-\delta_{jk}-\delta_{ki} + \delta_{ij}\delta_{jk} + \delta_{jk}\delta_{ki} + \delta_{ki}\delta_{ij} - \delta_{ij}\delta_{jk}\delta_{ki}) \nonumber \\+\delta_{il}\delta_{jn}\delta_{km}\left(1-\delta_{jk}\right)  \nonumber \\+
     \delta_{in}\delta_{jm}\delta_{kl}\left(1-\delta_{ki}\right)  \nonumber \\ +
     \delta_{im}\delta_{jl}\delta_{kn}\left(1-\delta_{ij}\right) \nonumber 
 \end{gather}
when contracted with the six-fold sum \ref{app:eq:G3General}. 

\subsubsection{Classical Light}
Consider once more our scenario where classical light is emitted 
 \begin{equation}
     \ket{\{\alpha\}} = \prod_\lambda^\nu \ket{\alpha_\lambda}
 \end{equation}
As with the pair correlation operator above, the triple correlation operator is diagonal in the coherent state basis. Proceeding analogously and defining 
 \begin{align} & \vec{q}_1 = \vec{k}_1 - \vec{k}_2 \\  & \vec{q}_2 = \vec{k}_2 - \vec{k}_3 \\  & \vec{q}_3 = \vec{k}_3 - \vec{k}_1 = -\vec{q}_1 - \vec{q}_2 
 \end{align}
we soon reach the expression
 \begin{multline}
     N \left\{ G^{(3)}(\vec{k}_1, \vec{k}_2, \vec{k}_3) \right\} \approx \left(\nu^3 - 3\nu^2 + 4\nu \right)\abs{\alpha}^6 \\ + \left(\nu - 2\right) \left( \abs{\sum_j^\nu \abs{\alpha}^3 e^{\iu \vec{q}_1 \cdot \vec{r}_j} }^2 + \abs{\sum_j^\nu \abs{\alpha}^3 e^{\iu \vec{q}_2 \cdot \vec{r}_j} }^2 + \abs{\sum_j^\nu \abs{\alpha}^3 e^{\iu \vec{q}_3 \cdot \vec{r}_j} }^2 \right) \\
     + \abs{\alpha}^6 \sum_{ijk} e^{\iu \vec{q}_1 \cdot \vec{r}_i} e^{\iu \vec{q}_2 \cdot \vec{r}_j} e^{\iu \vec{q}_3 \cdot \vec{r}_k} 
     + \abs{\alpha}^6 \sum_{ijk} e^{-\iu \vec{q}_1 \cdot \vec{r}_i} e^{-\iu \vec{q}_2 \cdot \vec{r}_j} e^{-\iu \vec{q}_3 \cdot \vec{r}_k}
 \end{multline}
remembering the over-counting rules from the previous section.
Observe that the last two terms are products of independent sums and complex conjugates. Moreover, each of the second, third, and fourth terms is a first-order correlation function \ref{app:eq:G1}. After normalizing by the cube of $\abs{\alpha}^2 \nu$, we write
 \begin{multline}
     \left\{g^{(3)}(\vec{k}_1, \vec{k}_2, \vec{k}_3) \right\} \approx \left(1 - \frac{3}{\nu} + \frac{4}{\nu^2} \right) + \left(1 - \frac{2}{\nu}\right) \left( \abs{g^{(1)}(\vec{q}_1) }^2 + \abs{g^{(1)}(\vec{q}_2) }^2 + \abs{g^{(1)}(\vec{q}_3) }^2 \right) \\
     + 2\text{Re}\left(g^{(1)}(\vec{q}_1) g^{(1)}(\vec{q}_2) g^{(1)}(\vec{q}_3) \right)
 \end{multline}
The last term here is the subject of the main text and allows retrieval of the exact structural phase $\phi(\vec{q})$ sought in coherent diffraction experiments.

\subsubsection{Quantum Light}
As with the pair correlations of quantum light, the triple correlator acting on non-classical states of the field picks up additional corrections during diagonalization. There are six terms (neglecting the over-counting correction terms, which must also be diagonalized) that must be diagonalized in the Fock basis. As an example, diagonalizing the first term of \ref{app:eq:G3Tensor} produces the operator 
 \begin{multline}
     \langle \hat{a}_i^\dag \hat{a}_j^\dag \hat{a}_k^\dag \hat{a}_i \hat{a}_j \hat{a}_k \rangle = 
     \langle \hat{a}_i^\dag \hat{a}_i \hat{a}_j^\dag \hat{a}_j \hat{a}_k^\dag \hat{a}_k - \hat{a}_i^\dag \hat{a}_i \hat{a}_j^\dag \hat{a}_k \delta_{jk} - \hat{a}_i^\dag \hat{a}_j \hat{a}_k^\dag \hat{a}_k \delta_{ij} + \hat{a}_i^\dag \hat{a}_k \delta_{ij} \delta_{kj} \\ - \hat{a}_i^\dag \hat{a}_k \hat{a}_j^\dag \hat{a}_j \delta_{ik} + \hat{a}_i^\dag \hat{a}_j \delta_{ik} \delta_{jk} \rangle
 \end{multline}
Clearly, obtaining the exact form of the triple correlations for chaotic quantum light is a tedious task. The result is not important for this paper, but it is important to realize that the classical and quantum cases lead to different expressions. We will skip this derivation for now.

\subsection{Friedel's Law and Symmetries of $\Phi(\vec{m}, \vec{n})$ \label{app:Symmetries}}
Given a real function $f(x)$, its Fourier Transform $F(k) = \mathcal{F}(f(x))$ has the following properties:
\begin{equation}
    F(k) = F^*(-k)
\end{equation}
\begin{equation}
    |F(k)|^2 = |F(-k)|^2
\end{equation}
For $\phi(k) = \text{arg}(F(k))$, 
\begin{equation}
    \phi(-k) = -\phi(k)
\end{equation}
leads to some useful symmetries of the correlation functions and phase map
\begin{equation}
    g^{(1)}(x) = g^{(1)*}(-x)
\end{equation}
\begin{equation}
     \Phi(m,n)=\Phi(n,m)
 \end{equation}
 \begin{equation}
     \Phi(0,n) = \Phi(m,0) = 0
 \end{equation}
 \begin{equation}
     \Phi(m,n) = \Phi(-n,-m)
 \end{equation}
 \begin{equation}
     \Phi(-m,m) = 0
 \end{equation}

\subsection{Coherent Diffraction Pattern Retrieval}
\begin{figure*}[htb]
\centering\includegraphics[width=\textwidth]{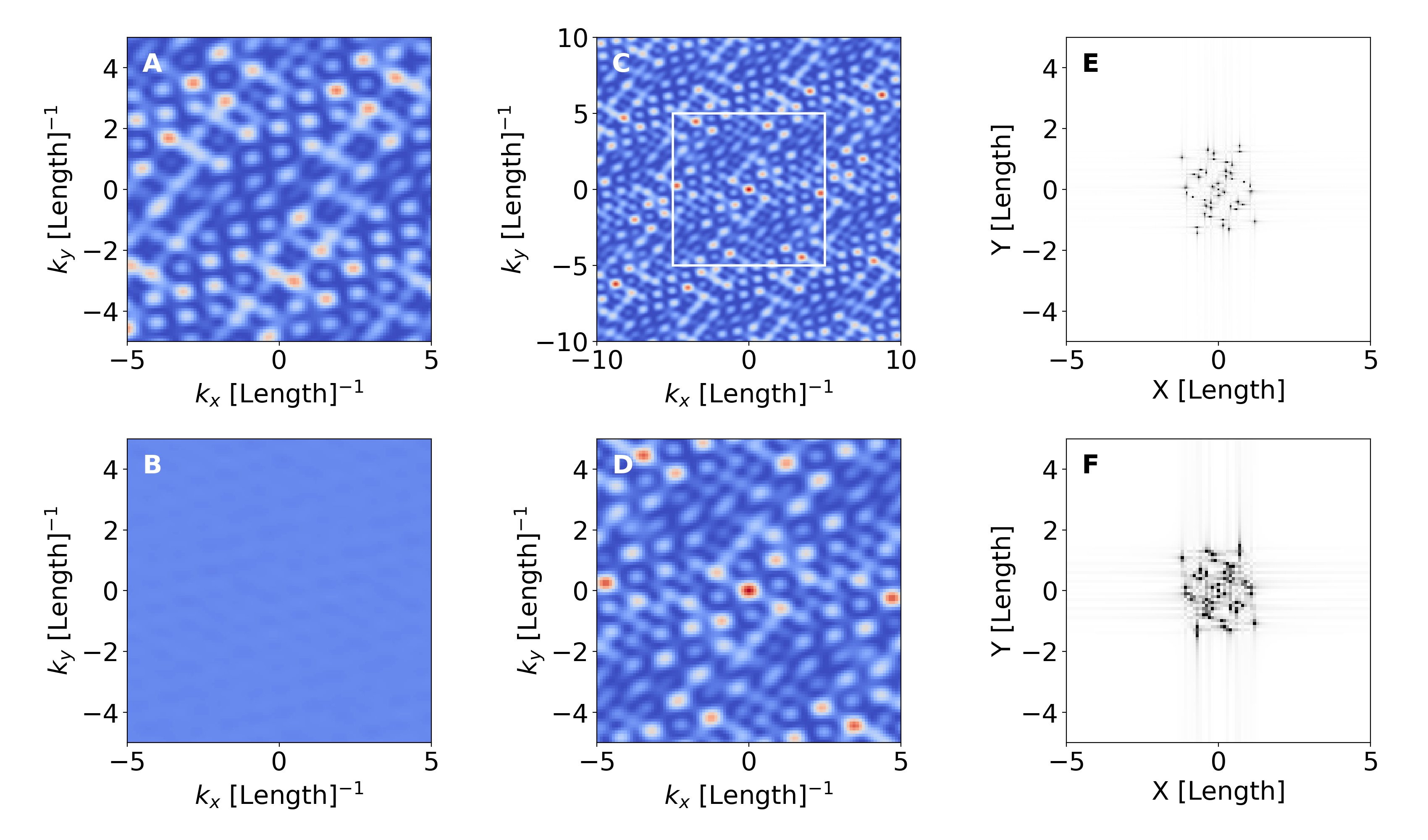}
\caption{Simulated fluorescence from a 2D array of seven classical point sources. (A) shows a speckle pattern, representing a single (random) phase relationship between point sources. (B) shows the mean of 1000 speckle patterns with the atom phases changing between shots. The result has nearly zero contrast. The sum of second-order (auto)correlations of the same shots (C) contains the coherent diffraction pattern (D) in the boxed region of (C) plus extra coherent diffraction data beyond the maximum scattering angle covered by the detector. The augmented $\vec{k}$-space window produces a real space (object) autocorrelation function with enhanced spatial resolution (E) compared to the real space autocorrelation (F) obtained from inverse Fourier inversion of (C) and (D).
\label{fig:SimulatedAmplitude}}
\end{figure*}

Fig.~\ref{fig:SimulatedAmplitude} shows the result of calculating the pairwise correlation function $g^{(2)}(\vec{q})$ from first-order incoherent fluorescence of atoms. The autocorrelation enables accurate retrieval of the coherent diffraction pattern out to $2|\vec{k}_{\text{max}}|$ (past the physical edge of the detector at $|\vec{k}_{\text{max}}|$) \cite{Trost_2020}.

\subsection{Extended Momentum Space Sampling Enhances Real Space Resolution}
\begin{figure*}[htb]
\centering\includegraphics[width=\textwidth]{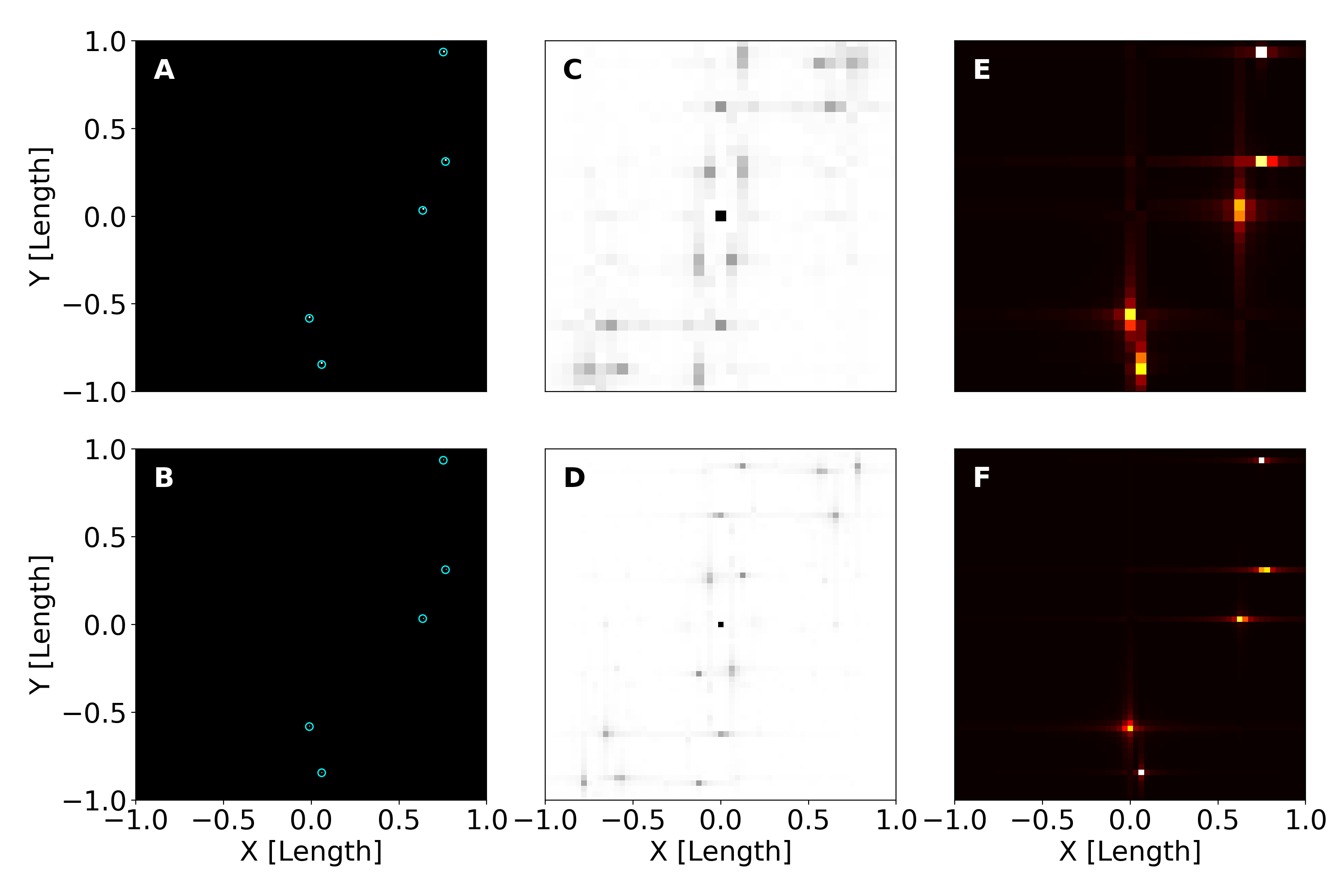}
\caption{(A) and (B) show the same array of atoms. (B) has double the spatial resolution (localization) of (A). The inverse Fourier Transform of simulated coherent diffraction (C) and (D) retrieves only the real space autocorrelation of the array since the phase information has been lost. In (E) and (F), the phase information is added back to break the translational symmetry and arrive at the original array. Note that in (D) and (F), where the simulated coherent diffraction goes out to $2\abs{\vec{k}_{\text{max}}}$, it is possible to resolve nearby atoms with much greater visibility than in (C) and (E), in which only data out to $\abs{\vec{k}_{\text{max}}}$ in momentum space is used.
\label{fig:ResolutionDemo}}
\end{figure*}

Fig.~\ref{fig:ResolutionDemo} shows that extended sampling of momentum space enhances the resolution of the real space autocorrelation and, if the phase information is available, the resolution of inverse Fourier Transform (the object).

\subsection{The Cross-Correlation Theorem}
For correlations of one-dimensional or small two-dimensional pixel arrays, computationally, it is usually faster to work directly with the outer product of the arrays, sum the resulting matrix along the diagonal, and divide out the underlying support (which represents correlations of unbunched photons). However, for larger arrays, this computation rapidly becomes memory-limited; for example, the second-order correlation of a typical 1000 by 1000 pixel array involves the calculation of a four-dimensional array represented by $1000^4$ floating point numbers. If each floating point number were 32 bits, we would need about 4 terabytes to store this array. 

The Cross-Correlation Theorem provides a simple method to calculate correlations using the Fast Fourier Transform. Let $f \bigstar g$ denote the cross correlation of $f(x)$ and $g(x)$. The Cross-Correlation Theorem states
\begin{equation}
    f \bigstar g = \mathcal{F}^{-1}[F(k) G^*(k)]
\end{equation} where $\mathcal{F}^{-1}$ indicates the inverse Fourier transform and $F(k)$ and $G(k)$ are the Fourier transforms of $f(x)$ and $g(x)$. If $f$ and $g$ are real-valued then
\begin{equation}
    f \bigstar g = \mathcal{F}^{-1}[F(k) G(-k)] = f(x) * g(-x)
\end{equation}
The autocorrelation $G^{(2)}(k_1,k_2)$ of intensities $I(k)$ on a detector is then 
\begin{equation}
    G^{(2)}(k_1-k_2) = I(k_1) \bigstar I(k_2) = I(k_1)*I(-k_2) 
\end{equation} where the last expression is the convolution of intensities.

The triple correlation
\begin{equation}
    G^{(3)}(k_1,k_2,k_3) = I(k_1) \bigstar I(k_2) \bigstar I(k_3)
\end{equation}
is most easily calculated from the bispectrum as 
\begin{equation}
    G^{(3)}(q_1, q_2) = \mathcal{F}^{-1}[\Tilde{I}(u)\Tilde{I}(v)\Tilde{I}(-u-v)]
\end{equation} where $u=k_1-k_2$ and $v=k_2-k_3$.

\subsection{Coarse Binning and Interpolation of Phase Information}
A major difficulty of performing a triple correlation experiment is the computational complexity of the method. For a square, two-dimensional detector with side length of $N$ pixels, the third-order correlation function (correlations between triples of pixels) is a six-dimensional function. This must either be stored in memory or rapidly generated on-the-fly during phase retrieval, neither of which is an attractive option for working with large detectors.

\begin{figure}[htb]
\centering\includegraphics[width=\textwidth]{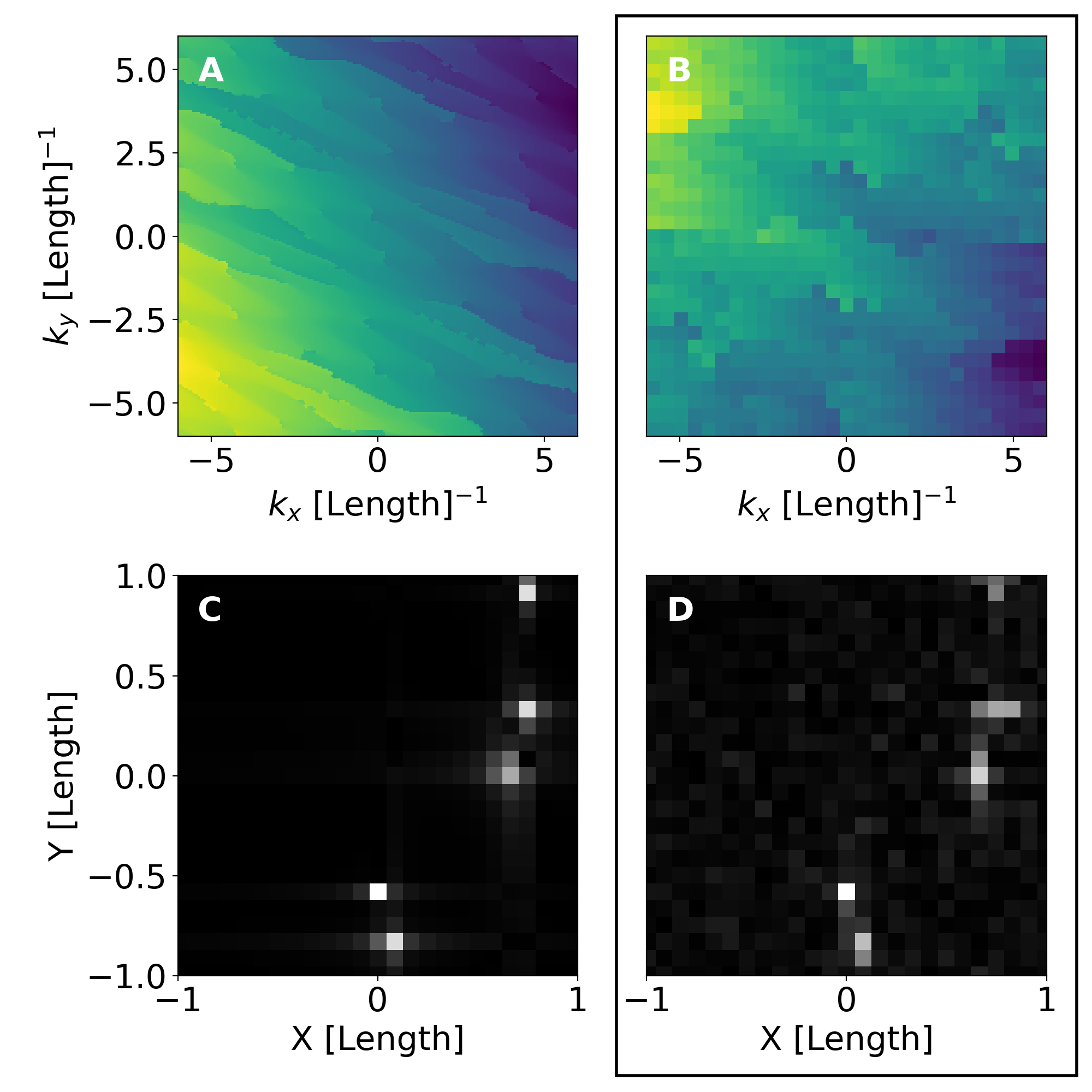}
\caption{(A) shows the phase $\phi(\vec{k})$ sampled and unwrapped across 201 by 201 pixels. (B) shows $\phi(\vec{k})$ sampled and unwrapped across 29 by 29 pixels on the same area of $\vec{k}$-space. (C) is the result of taking the inverse Fourier transform using the phase shown in (A). (D) is the result of taking the inverse Fourier transform using the phase in (B) (interpolated to the same sampling resolution as in (A)) and the same modulus used to produce (C). (D) suffers from additional noise compared to (C), but each of the five atoms remains well-resolved.
\label{fig:CoarsePhase}}
\end{figure}

We can consider a case where the sampling rate of the triple correlation may be reduced, the phase retrieved and interpolated, and combined with densely sampled diffraction data. The upscaled phase information breaks the symmetry of the real space structure autocorrelation, selecting peaks at the true atomic positions. For example, Fig.~\ref{fig:CoarsePhase} shows that phase information sampled at 30 pixels combined with the modulus from coherent diffraction data sampled at 200 pixels across the same region of $\vec{k}$-space retrieves a noisy, but discernible structure.

\subsection{Linear Phase Ramps and Sign Flips Shift and Invert the Fourier Transform \label{app:PhaseRamp}}
\begin{figure*}[htb]
\centering\includegraphics[width=\textwidth]{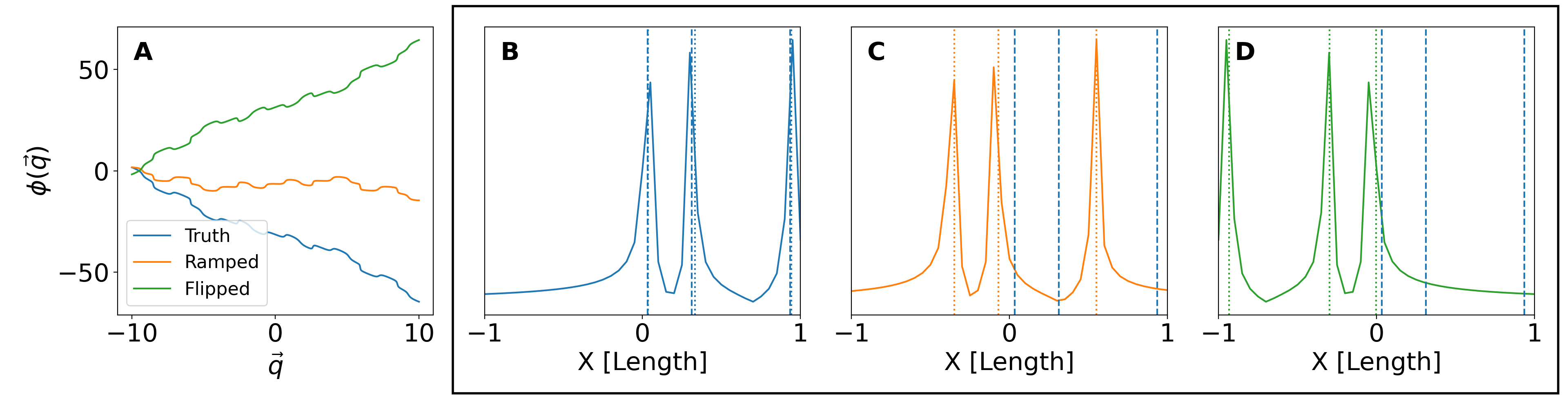}
\caption{(A) shows three phases which, combined with identical modulus data, produce different inverse Fourier transforms in (B), (C), and (D). The blue phase (A) and Fourier transform (B) represent the truth values. The orange phase (A) has acquired an additional linear ramp compared to the truth phase which results in a shifted Fourier transform (C) compared to the truth Fourier transform in (B). The green phase (A) has a sign flip with respect to the truth phase, causing the Fourier transform in (D) to appear inverted across the origin. Compounded ramps and sign flips of the phase result in compounded shifts and inversions of the Fourier transform. In each of (B), (C), and (D), the blue vertical dashes mark the truth positions of the atoms used in the model. The vertical dots (color matches color of phase pictured in (A)) indicate the atom positions calculated via harmonic inversion of the modulus and phase. In the case of (B), harmonic inversion determines the truth positions of the atoms quite accurately. In (C) and (D), sub-linewidth positions of the shifted and inverted atom array are shown to maintain the correct interatomic separations. The Fourier transforms (solid curves) are each normalized to the height of the tallest peak.
\label{fig:PhaseRamp}}
\end{figure*}

Fig.~\ref{fig:PhaseRamp} demonstrates the effect of linear ramps and sign flips of the phase in real space. A linear phase ramp in momentum space produces a translation in real space. A sign flip of the entire phase inverts the real space structure. Both exact numerical solving and differential evolution structure fitting obtain the phase accurately up to a linear phase ramp or sign flip (or both) in most cases, maintaining the correct relative positions but producing incorrect absolute positions and orientations.

\subsection{Harmonic Inversion Resolves Real Space Positions from Fourier Data \label{app:HarmonicInvert}}
In Fig.~\ref{fig:PhaseRamp} we used harmonic inversion to retrieve precise real space positions of the atoms from the Fourier modulus and phase. The inverse Fourier transform gives peaks with a finite width and, thus, uncertainty in the position of the atom. Harmonic inversion is not limited by the Fourier uncertainty principle and superresolves the positions of the atoms directly from the momentum space data rather than the Fourier transform \cite{Mandelshtam_1997, Mandelshtam_1998, Mandelshtam_2003}.

\subsection{1D Phase Retrieval Example}


\subsubsection{Exact Coherent Phase Retrieval}
Using Equation \ref{eq:TriplePhase} and the algorithm described in \ref{sect:Algorithm}, we can calculate the Fourier phase from triple correlation data and compare the result to the true value. Figure~\ref{fig:PhaseRetrieval_1D} shows the results of phase retrieval in a simulation with a 1D pixel detector.

\subsubsection{Coherent Phase Retrieval via Fitting}
Retrieval of the Fourier phase is highly sensitive to statistical and systematic noise. The examples of exact phase retrieval shown in the previous section were calculated for ideal experimental conditions with no sources of noise. Noise is especially problematic in the division step of Equation \ref{eq:TriplePhase}. Moreover, it is expected that the phase noise will become problematic for sources with large numbers of incoherent atoms \cite{Trost_2020}. Therefore, phase retrieval via an optimization algorithm which compares the closure phase (Equation \ref{eq:PhiDef}) of a trial atom array to a measured closure phase may be more tractable for real experimental data.

\begin{figure*}[htb]
\centering\includegraphics[width=\textwidth]{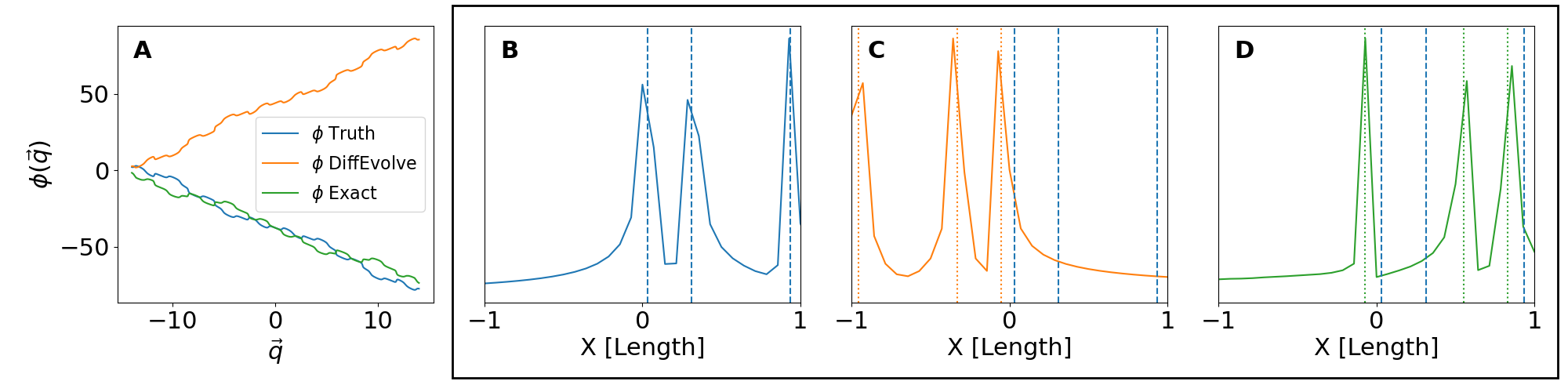}
\caption{Phase retrieval results compared to truth values (blue) for a 1D array of three atoms using the exact numerical solution (green) of Section \ref{sect:Algorithm} and differential evolution (orange). (A) shows the retrieved (orange and green) and truth (blue) phases used, together with the modulus obtained from Equation \ref{eq:SiegertModified}, to execute the inverse Fourier transform in (B), (C), and (D). In (B), the truth object from the inverse Fourier transform and the known positions (vertical dots) of the atoms match, unsurprisingly. (C) shows that differential evolution correctly found atom positions (orange vertical dots) with the correct separation and ordering, but the structure appears flipped and shifted compared to the truth positions (blue vertical dots). This is due to the linear phase ramp in (A) compared to the truth phase. The effect of linear phase ramps and sign flips is detailed in Supplement \ref{app:PhaseRamp}. Similarly, the exact numerical solution in (D) also retrieved the correct interatomic separations but the phase (green) in (A) is inverted to the truth phase, causing a flip of the real space structure. Since the exact numerical algorithm finds the phase without directly determining atom positions, we used harmonic inversion (see Supplement \ref{app:HarmonicInvert}) to determine a sub-linewidth position for the three atoms (green vertical dots) to compare to the truth atom positions (blue vertical dots). The Fourier transforms (solid curves) are each normalized to the height of the tallest peak.
\label{fig:PhaseRetrieval_1D}}
\end{figure*}

We used a differential evolution algorithm to iterate on the spatial arrangement of the atoms until satisfactory convergence between the simulated closure phase and the closure phase of the trial structure was achieved. It can be seen in Fig.~\ref{fig:PhaseRetrieval_1D} that the exact method based on the algorithm and the structure optimization method achieve similar results.

\subsection{Alternate Phase Toggling}
\begin{figure*}[htb]
\centering\includegraphics[width=\textwidth]{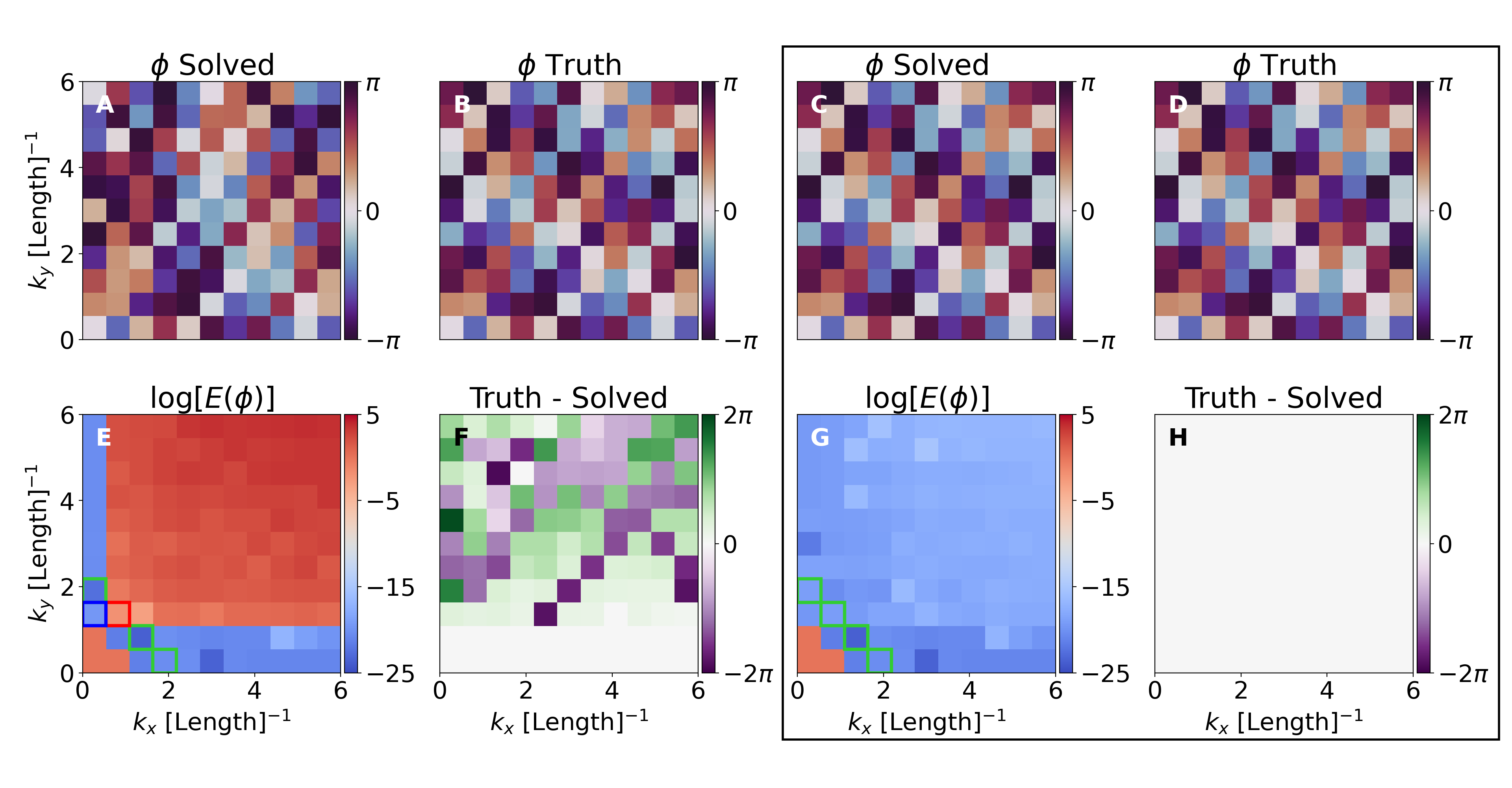}
\caption{In these examples, the value of $\phi(\vec{k})$ in (A) and (C) in each diagonal $k_y = -k_x + n$, $n \in \mathbb{Z}$ is determined beginning at the origin and ending in the upper right corner. In the first example (left), where toggling is not used, the solved value of $\phi$ in (A) has a large difference (F) from the true value of $\phi$ in (B). We observe that the value of $\log\left[ E(\phi) \right]$ (E) is seen to spike (red box) along the third diagonal (green and red boxes). The value of the error function increases and propagates to surrounding pixels as additional diagonals are solved. This indicates that a pixel in the second diagonal (blue box) has been assigned incorrectly. Resolving the second diagonal by substituting alternate minima of $\log\left[ E(\phi) \right]$ eventually arrives at a $\phi(\vec{k})$ (C) which has uniformly low values of the error function (G) such that the difference (H) between the solution (C) and truth (D) is quite small. Note that the error function, (E) and (G), is defined to be zero at the origin and the origin nearest neighbors as these values of $\phi(\vec{k})$ are free parameters of the solution process.
\label{fig:PhaseTogglingAppendix}}
\end{figure*}

Figure~\ref{fig:PhaseTogglingAppendix} illustrates how phase toggling is implemented in the algorithm from the main text.

\end{document}